\let\ifarxiv=\iftrue     
\pdfoutput=1
{\def\usepackage{ws-procs9x6}}
\documentclass[12pt,a4paper]{article}

\ifarxiv\ifnum\pdfoutput=1\else
\PassOptionsToPackage{hypertex}{hyperref}
\PassOptionsToPackage{draft}{graphicx}
\usepackage{showkeys}
\fi\fi

\usepackage{amsmath,amssymb}
\usepackage[bookmarks=true,hyperfigures=true]{hyperref}
\usepackage{graphicx}
\usepackage[nosort]{cite}
\usepackage[bulletsep]{collref}

\usepackage[a4paper,text={450pt,650pt},centering]{geometry}

\let\oldbfseries=\bfseries
\let\oldmdseries=\mdseries
\let\oldnormalfont=\normalfont
\renewcommand{\bfseries}{\oldbfseries\boldmath}
\renewcommand{\mdseries}{\oldmdseries\unboldmath}
\renewcommand{\normalfont}{\oldnormalfont\unboldmath}

\allowdisplaybreaks[3]

\numberwithin{equation}{section}

\usepackage[font=small,labelfont=bf,width=0.85\textwidth]{caption}

\providecommand{\hypersetup}[1]{}
\providecommand{\texorpdfstring}[2]{#1}
\providecommand{\pdfbookmark}[3][]{}

\hypersetup{plainpages=false}
\hypersetup{pdfpagemode=UseNone}
\hypersetup{bookmarksnumbered=true}
\hypersetup{pdfstartview=FitH}
\hypersetup{colorlinks=false}
\hypersetup{citebordercolor={.5 1 .5}}
\hypersetup{urlbordercolor={.5 1 1}}
\hypersetup{linkbordercolor={1 .7 .7}}

\makeatletter
\let\@keywords\@empty
\let\@subject\@empty
\providecommand{\keywords}[1]{\gdef\@keywords{#1}}
\providecommand{\subject}[1]{\gdef\@subject{#1}}
\def\thetitle{\@title}
\def\theauthor{\@author}
\def\thesubject{\@subject}
\def\thedate{\@date}
\def\thekeywords{\@keywords}
\makeatother
\AtBeginDocument{
\hypersetup{pdftitle={\thetitle}}%
\hypersetup{pdfauthor={\theauthor}}%
\hypersetup{pdfsubject={\thesubject}}%
\hypersetup{pdfkeywords={\thekeywords}}%
}

\DeclareMathSymbol{\Gamma}{\mathalpha}{letters}{"00}
\DeclareMathSymbol{\Delta}{\mathalpha}{letters}{"01}
\DeclareMathSymbol{\Theta}{\mathalpha}{letters}{"02}
\DeclareMathSymbol{\Lambda}{\mathalpha}{letters}{"03}
\DeclareMathSymbol{\Xi}{\mathalpha}{letters}{"04}
\DeclareMathSymbol{\Pi}{\mathalpha}{letters}{"05}
\DeclareMathSymbol{\Sigma}{\mathalpha}{letters}{"06}
\DeclareMathSymbol{\Upsilon}{\mathalpha}{letters}{"07}
\DeclareMathSymbol{\Phi}{\mathalpha}{letters}{"08}
\DeclareMathSymbol{\Psi}{\mathalpha}{letters}{"09}
\DeclareMathSymbol{\Omega}{\mathalpha}{letters}{"0A}

\newcommand{\gen}[1]{\mathrm{#1}}
\newcommand{\gencen}{\gen{A}}
\newcommand{\gender}{\gen{B}}
\newcommand{\genaffcen}{\gen{C}}
\newcommand{\genaffder}{\gen{D}}

\newcommand{\copro}{\mathrm{\Delta}}
\newcommand{\coproop}{\widetilde{\copro}}

\newcommand{\rmat}{\mathcal{R}}

\newcommand{\Tr}{\mathop{\mathrm{Tr}}}

\newcommand{\order}[1]{\mathcal{O}(#1)}

\newcommand{\Integers}{\mathbb{Z}}

\ifx\genfrac\sdflkaj\else\fi
\newcommand{\sfrac}[2]{{\textstyle\frac{#1}{#2}}}
\newcommand{\half}{\sfrac{1}{2}}
\newcommand{\ihalf}{\sfrac{i}{2}}
\newcommand{\quarter}{\sfrac{1}{4}}

\newcommand{\matr}[2]{\left(\begin{array}{#1}#2\end{array}\right)}
\newcommand{\trans}{{\scriptscriptstyle\mathrm{T}}}
\newcommand{\strans}{{\scriptscriptstyle\mathrm{ST}}}

\newcommand{\lrbrk}[1]{\left(#1\right)}
\newcommand{\bigbrk}[1]{\bigl(#1\bigr)}

\newcommand{\bigcomm}[2]{\big[#1,#2\big]}
\newcommand{\comm}[2]{[#1,#2]}

\newcommand{\acomm}[2]{\{#1,#2\}}
\newcommand{\bigacomm}[2]{\big\{#1,#2\big\}}

\newcommand{\state}[1]{\mathopen{|}#1\mathclose{\rangle}}

\newcommand{\spn}[1]{\langle #1\rangle}
\newcommand{\cybe}[2]{[[#1,#2]]}

\newcommand{\xp}[1]{x^{+}_{#1}}
\newcommand{\xm}[1]{x^{-}_{#1}}
\newcommand{\xpm}[1]{x^{\pm}_{#1}}

\newcommand{\conw}{}
\newcommand{\coni}{}

\newcommand{\alg}[1]{\mathfrak{#1}}

\newcommand{\grp}[1]{\mathrm{#1}}

\newcommand{\nn}{\nonumber}
\newcommand{\nln}{\nonumber\\}
\newcommand{\nl}[1][0pt]{\nonumber\\[#1]&\hspace{-4\arraycolsep}&\mathord{}}

\newcommand{\earel}[1]{\mathrel{}&\hspace{-2\arraycolsep}#1\hspace{-2\arraycolsep}&\mathrel{}}
\newcommand{\eq}{\earel{=}}
\newcommand{\beq}{\begin{equation}}
\newcommand{\eeq}{\end{equation}}

\def\[{\begin{equation}}
\def\]{\end{equation}}
\def\<{\begin{eqnarray}}
\def\>{\end{eqnarray}}

\makeatletter
\def\mr@ignsp#1 {\ifx\:#1\@empty\else #1\expandafter\mr@ignsp\fi}%
\newcommand{\multiref}[1]{\begingroup
\xdef\mr@no@sparg{\expandafter\mr@ignsp#1 \: }%
\def\mr@comma{}%
\@for\mr@refs:=\mr@no@sparg\do{\mr@comma\def\mr@comma{,}\ref{\mr@refs}}%
\endgroup}
\makeatother

\newcommand{\hypref}[2]{\ifx\href\asklfhas #2\else\href{#1}{#2}\fi}

\newcommand{\secref}[1]{Sec.~\multiref{#1}}

\newcommand{\tabref}[1]{Tab.~\multiref{#1}}

\newcommand{\figref}[1]{Fig.~\multiref{#1}}
\renewcommand{\eqref}[1]{(\multiref{#1})}

\makeatletter
\newlength{\apb@width}
\newcommand{\autoparbox}[2][c]{\settowidth{\apb@width}{#2}\parbox[#1]{\apb@width}{#2}}
\newcommand{\includegraphicsbox}[2][]{\autoparbox{\includegraphics[#1]{#2}}}
\makeatother

\ifx\href\asklfhas\newcommand{\href}[2]{#2}\fi
\newcommand{\arxivlink}[1]{\href{http://arxiv.org/abs/#1}{arxiv:#1}}

\title{The Classical Trigonometric r-Matrix for\texorpdfstring{\\}{ }the Quantum-Deformed Hubbard Chain}
\author{Niklas Beisert}
\subject{MSC (2010): 17B62, 81U15, 17B67, 17B25}

\begin{document}

\pdfbookmark[1]{Title Page}{title}

\renewcommand{\thefootnote}{\fnsymbol{footnote}}
\thispagestyle{empty}
\begin{flushright}\footnotesize
\texttt{\arxivlink{1002.1097}}\\
\texttt{AEI-2010-016}
\end{flushright}
\vspace{1cm}

\begin{center}%
\begingroup\Large\bfseries%
\thetitle%
\ifarxiv\footnote{Typeset in $\mathfrak{Go}$\TeX-Lite$^{\mathfrak{tm}}$}\fi%
\par\endgroup\vspace{1cm}%

\textsc{\theauthor}\vspace{5mm}%
\hypersetup{pdfauthor={Niklas Beisert}}%

\textit{Max-Planck-Institut f\"ur Gravitationsphysik\\%
Albert-Einstein-Institut\\%
Am M\"uhlenberg 1, 14476 Potsdam, Germany}\vspace{3mm}%

\texttt{nbeisert@aei.mpg.de}%
\par
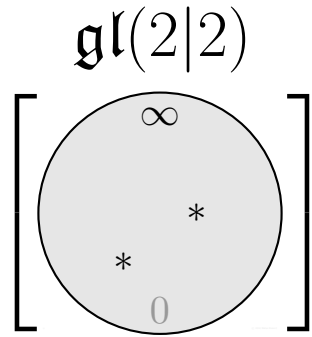
\vspace{1cm}

\ifarxiv%
{\huge$\alg{gl}(2|2)$\par}\vspace{0.2cm}
{\huge$\phantom{\alg{gl}(2|2)}\left[\includegraphicsbox{FigSphereTitle.mps}\right]\phantom{\alg{gl}(2|2)}$}
\par\vspace{1cm}
\else\vspace{3cm}
\fi

\textbf{Abstract}\vspace{7mm}

\begin{minipage}{12.7cm}
The one-dimensional Hubbard model is an exceptional integrable spin chain
which is apparently based on a deformation of the Yangian for the 
superalgebra $\alg{gl}(2|2)$.
Here we investigate the quantum-deformation of the Hubbard model
in the classical limit. 
This leads to a novel classical r-matrix of trigonometric kind. 
We derive the corresponding one-parameter family of Lie bialgebras as a 
deformation of the affine $\alg{gl}(2|2)$ Kac--Moody superalgebra. 
In particular, we discuss the affine extension as well as discrete symmetries,
and we scan for simpler limiting cases, 
such as the rational r-matrix for the undeformed Hubbard model.
\ifarxiv\else\par\vspace{1.5ex}
\textbf{MSC Classes:}
17B62, 
81U15, 
17B67, 
17B25. 
\par\vspace{1.5ex}
\fi
\end{minipage}

\end{center}

\newpage
\renewcommand{\thefootnote}{\arabic{footnote}}
\setcounter{footnote}{0}


\section{Introduction and Overview}

The Hubbard model \cite{Hubbard:1963aa}
is a model of spin-half electrons
hopping around on a lattice of atoms
(see \cite{Essler:2005aa} for an introduction). 
It has several useful features that make it attractive
for the investigation of aspects of electron transport,
in particular superconductivity.
An unrelated property of its one-dimensional incarnation
is \emph{integrability} which enabled 
Lieb and Wu to find the spectrum 
by means of Bethe equations \cite{Lieb:1968aa}.
Remarkably, the integrable structure is different from 
conventional spin chain models in several respects:
The most striking distinction is, arguably, 
that the R-matrix, which was found by Shastry \cite{Shastry:1986bb}, 
is not of difference form.%
\footnote{Two similar cases have previously been discussed:
These are based on the twisted affine superalgebras 
$\alg{gl}(N|N)^{(2\prime)}$ \cite{Nazarov:1999aa}
and $\alg{d}(2,1;e^{2\pi i/3})^{(3)}$ \cite{Leites:1984aa}.
Their Cartan--Killing forms are charged under the twisting automorphism
which leads to unconventional quantum algebras. 
Another exceptional case involving the twisted affine superalgebra
$\alg{gl}(2|2)^{(2)}$ is discussed in \protect\secref{sec:rattwist}.}
This implies that the description of the integrable structure through 
standard \emph{Yangian} or \emph{quantum affine algebras} \cite{Drinfeld:1985rx,Drinfeld:1986in,Jimbo:1985zk,Jimbo:1985ua}
cannot apply to this case.%
\footnote{The R-matrix must be invariant under the affine shift 
which enforces the difference form.}

For a long time the question of the algebraic structure underlying 
the Hubbard chain was left at rest. 
Recent progress towards this goal came from 
a totally unexpected direction:
It turned out that Shastry's R-matrix 
is equivalent \cite{Beisert:2006qh} to a 
scattering matrix \cite{Beisert:2005tm,Staudacher:2004tk}
found in the context of the AdS/CFT correspondence \cite{Maldacena:1998re} 
(see \cite{Beisert:2010jr,Beisert:2004ry,Plefka:2005bk,Arutyunov:2009ga} for reviews
of integrability in AdS/CFT).
This matrix has a \emph{centrally extended} $\alg{psl}(2|2)$ supersymmetry by construction
which includes the two (more or less) manifest $\alg{sl}(2)$ symmetries 
of the Hubbard model \cite{Lieb:1989aa,Yang:1989aa}.
Since then, there has been a lot of progress 
in the formulation of a quantum symmetry algebra
for Shastry's R-matrix
\cite{Gomez:2006va,Plefka:2006ze,Beisert:2007ds,Matsumoto:2007rh,Spill:2008tp,Torrielli:2008wi,Spill:2008yr,Matsumoto:2009rf}.
In particular, the construction for higher representations
has advanced significantly
\cite{Beisert:2006qh,Chen:2006gp,Arutyunov:2008zt,deLeeuw:2008dp,deLeeuw:2008ye,Arutyunov:2009mi,Arutyunov:2009ce,Arutyunov:2009iq,Arutyunov:2009pw}.
Still, it is fair to say that a satisfactory quantisation 
to a \emph{quasi-triangular Hopf algebra} 
similar to a Yangian has not yet been achieved. 

\smallskip

By quantum-deforming%
\footnote{The q-deformation lifts a rational to a trigonometric R-matrix, 
e.g.\ Heisenberg XXX to XXZ.}
the Hubbard chain we hope to get further insights 
into the Hopf algebra underlying this special model:
For conventional integrable spin chains based on Lie (super)algebra symmetries, 
the quantum deformation lifts the Yangian to a quantum affine algebra.
This has some drawbacks, but also benefits.
One the one hand, the deformation breaks the manifest Lie symmetry
down to its Cartan subalgebra.
On the other hand, one gains a more uniform and symmetric description
of the algebra itself.
It is then possible to return to the undeformed model
and recover the Yangian as a particular limit.
The limit is singular, and it obscures some of the symmetry
of the quantum affine formulation.
An increased internal symmetry will hopefully simplify 
the formulation of a quasi-triangular Hopf algebra for the 
(quantum-deformed) Hubbard chain. 
Another motivation to study the quantum-deformation 
is that some of the structures in the centrally extended $\alg{psl}(2|2)$ algebra \cite{Beisert:2005tm} 
for Shastry's R-matrix are reminiscent of quantum affine algebras. 

The quantum-deformation of the Hubbard Hamiltonian 
along with its R-matrix was performed in \cite{Beisert:2008tw}. 
It has an additional parameter $q$, and therefore yields a bigger class of models. 
It turned out that this class contains a multi-parameter family of 
deformations of the Hubbard model proposed earlier 
by Alcaraz and Bariev \cite{Alcaraz:1999aa}.
In fact, many of the variants of the Hubbard chain 
(see references in \cite{Beisert:2006qh,Beisert:2008tw})
are special cases of this model.
The deformed and undeformed model and R-matrix
have in common a rather complicated structure
which obstructs direct attempts to set up a quasi-triangular Hopf algebra.

\medskip

Fortunately, there is a limit, the \emph{classical limit}, 
which makes the algebraic structure much more tractable:
The classical framework consists of some Lie algebra $\alg{g}$ along 
with an element $r$ of the tensor product $\alg{g}\otimes\alg{g}$ 
serving as the \emph{classical r-matrix}.
For the quantum algebra $\alg{g}$ is promoted to a
deformation of its universal enveloping algebra
$\grp{U}_q(\alg{g})$ 
which is substantially bigger than $\alg{g}$ itself.
For r-matrices with spectral parameter,
the Lie algebra $\alg{g}$ is typically of affine Kac--Moody type,
for which an efficient and uniform description exists. 
All in all, the manipulations in the classical limit can 
usually be performed very explicitly with pen and paper,
much in contradistinction to the quantum case.

The classical limit of Shastry's R-matrix 
was derived in \cite{Klose:2006zd,Torrielli:2007mc}. 
The underlying Lie algebra with universal classical r-matrix 
was found in \cite{Beisert:2007ty}. 
This algebra turned out to be a peculiar deformation of the 
loop algebra $\alg{gl}(2|2)[u,u^{-1}]$. 
Note that the $\alg{gl}(2|2)$ algebra is not simple, it contains central charges 
as well as derivations \cite{Moriyama:2007jt,Matsumoto:2007rh,Beisert:2007ty},
and thus it escapes the classification of r-matrices
in \cite{Leites:1984aa}.
The algebra is curious because it is not a loop algebra of some deformed algebra, 
the deformation applies to the loop algebra structure itself,
in particular to the derivations and charges.
Yet, surprisingly, the algebra admits a quasi-triangular bialgebra structure.

\medskip

In this paper we will derive the classical r-matrix 
for the quantum-deformed Hubbard chain. 
This is the trigonometric analog of the 
rational r-matrix in \cite{Klose:2006zd,Torrielli:2007mc,Beisert:2007ty}.
We expect that it will be of help in deriving the 
full quantum algebra framework for the
(quantum-deformed) Hubbard model.

The paper is organised as follows:
We start with a brief review of the quantum R-matrix in \secref{sec:quantum}.
In the following \secref{sec:classical} we perform the classical limit 
and show that it leads to a quasi-triangular Lie bialgebra.
Next we consider its affine extension in \secref{sec:affine}
which provides some more structure to the algebra. 
The r-matrix and the algebra have several discrete symmetries
and special points which are discussed in \secref{sec:discrete}.
The last \secref{sec:limits} is devoted to the enumeration of simpler limiting cases
of the r-matrix and the algebra. 
Finally, in \secref{sec:concl} we conclude and give an outlook.

\section{Quantum-Deformed S-Matrix}
\label{sec:quantum}

In \cite{Beisert:2008tw} a quantum-deformation of the 
centrally extended $\alg{psl}(2|2)$ algebra was defined.
Subsequently, the fundamental R-matrix for this algebra was derived.
In this section we will summarise the results of \cite{Beisert:2008tw}
important to this paper.

\subsection{Serre--Chevalley Presentation}

\begin{figure}\centering
\includegraphics{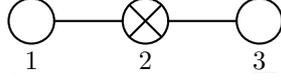}
\caption{Distinguished Dynkin diagram for $\alg{sl}(2|2)$.}
\label{fig:Dynkin}
\end{figure}

We first define the quantum deformation of the extended $\alg{psl}(2|2)$ algebra
in the Serre--Chevalley presentation,
cf.\ \cite{Scheunert:1993aa,Zhang:1999aa} for the case of conventional (affine) $\alg{gl}(2|2)$.
It has 9 Serre--Chevalley generators
$\gen{H}_j,\gen{E}_j,\gen{F}_j$ with $j=1,2,3$.
For the distinguished choice of Dynkin diagram of $\alg{psl}(2|2)$,
see \figref{fig:Dynkin},
the generators $\gen{E}_2,\gen{F}_2$ 
are fermionic while the remaining 7 are bosonic.
The symmetric Cartan matrix $A_{jk}$ reads%
\footnote{For superalgebras it is sometimes convenient 
to flip the signs of some rows/columns to make the matrix symmetric.}
\[
A_{jk}=\matr{rrr}{+2&-1&0\\-1&0&+1\\0&+1&-2}.
\]

\paragraph{Algebra.}

The commutators with symmetrised Cartan elements $\gen{H}_j$ 
are determined by the Cartan matrix $A_{jk}$
\[
\comm{\gen{H}_j}{\gen{H}_k}=0,
\qquad
\comm{\gen{H}_j}{\gen{E}_k}=+A_{jk}\gen{E}_k,
\qquad
\comm{\gen{H}_j}{\gen{F}_k}=-A_{jk}\gen{F}_k.
\]
The commutators between $\gen{E}_j$ and $\gen{F}_k$ 
are non-trivial only for $j=k$
\[\label{eq:EFcomm}
\comm{\gen{E}_1}{\gen{F}_1}=
\frac{q^{\gen{H}_1}-q^{-\gen{H}_1}}{q-q^{-1}}\,,
\qquad
\acomm{\gen{E}_2}{\gen{F}_2}=-
\frac{q^{\gen{H}_2}-q^{-\gen{H}_2}}{q-q^{-1}}\,,
\qquad
\comm{\gen{E}_3}{\gen{F}_3}=-
\frac{q^{\gen{H}_3}-q^{-\gen{H}_3}}{q-q^{-1}}\,.
\]
The Serre relations between 
alike generators $\gen{E}_j$ or $\gen{F}_j$
read
\< 0\eq
\comm{\gen{E}_1}{\gen{E}_3}
=\comm{\gen{F}_1}{\gen{F}_3} =\gen{E}_2\gen{E}_2
=\gen{F}_2\gen{F}_2
\\
\eq\gen{E}_1\gen{E}_1\gen{E}_2
-(q+q^{-1})\gen{E}_1\gen{E}_2\gen{E}_1
+\gen{E}_2\gen{E}_1\gen{E}_1
=\gen{E}_3\gen{E}_3\gen{E}_2
-(q+q^{-1})\gen{E}_3\gen{E}_2\gen{E}_3
+\gen{E}_2\gen{E}_3\gen{E}_3
\nln\nonumber
\eq\gen{F}_1\gen{F}_1\gen{F}_2
-(q+q^{-1})\gen{F}_1\gen{F}_2\gen{F}_1
+\gen{F}_2\gen{F}_1\gen{F}_1
=\gen{F}_3\gen{F}_3\gen{F}_2
-(q+q^{-1})\gen{F}_3\gen{F}_2\gen{F}_3
+\gen{F}_2\gen{F}_3\gen{F}_3
.
\>
%

\paragraph{Central Elements.}

What singles out $\alg{psl}(2|2)$ from the other simple
superalgebras is that it has three non-trivial central extensions
\cite{Nahm:1977tg,Iohara:2001aa}.
Our algebra has two central elements $\gen{C},\gen{D}$,
and they are the key to the peculiar features discussed in this paper.
The standard central element $\gen{C}$ in $\alg{sl}(2|2)$ reads
\[\label{eq:quantumC}
\gen{C}=-\half\gen{H}_1-\gen{H}_2-\half\gen{H}_3.
\]
In addition there are two exceptional central elements $\gen{P}$, $\gen{K}$
which originate from dropping the two Serre relations $\gen{P}=\gen{K}=0$
particular to superalgebras
\cite{Scheunert:1993aa}
\<
\label{eq:quantumPK}
\gen{P}\eq \gen{E}_1\gen{E}_2\gen{E}_3\gen{E}_2
+\gen{E}_2\gen{E}_3\gen{E}_2\gen{E}_1
+\gen{E}_3\gen{E}_2\gen{E}_1\gen{E}_2
+\gen{E}_2\gen{E}_1\gen{E}_2\gen{E}_3
-(q+q^{-1})\gen{E}_2\gen{E}_1\gen{E}_3\gen{E}_2,
\nln
\gen{K}\eq \gen{F}_1\gen{F}_2\gen{F}_3\gen{F}_2
+\gen{F}_2\gen{F}_3\gen{F}_2\gen{F}_1
+\gen{F}_3\gen{F}_2\gen{F}_1\gen{F}_2
+\gen{F}_2\gen{F}_1\gen{F}_2\gen{F}_3
-(q+q^{-1})\gen{F}_2\gen{F}_1\gen{F}_3\gen{F}_2.
\>
In order to get an interesting quantum algebra structure
the two extra central elements have to be constrained.
We introduce a new central element $\gen{D}$ 
as well as two global constants $g,\alpha$,
and express $\gen{P},\gen{K}$ through them%
\footnote{Notice the similarity between \eqref{eq:quantumPK,eq:PKident} and \eqref{eq:EFcomm}.} 
\[\label{eq:PKident}
\gen{P}=g\alpha(1-q^{2\gen{C}}q^{2\gen{D}}),\qquad
\gen{K}=g\alpha^{-1}(q^{-2\gen{C}}-q^{-2\gen{D}}).
\]

\paragraph{Coalgebra.}

The standard quantum-deformed coproduct applies to all
\emph{bosonic} generators $\gen{E}_j,\gen{F}_j,\gen{H}_j$ 
(i.e.\ all except $\gen{E}_2$ and $\gen{F}_2$)
\<\label{eq:qcoproduct}
\copro(\gen{H}_j)\eq \gen{H}_j\otimes 1+1\otimes \gen{H}_j, \nln
\copro(\gen{E}_j)\eq \gen{E}_j\otimes 1+q^{-\gen{H}_j}\otimes \gen{E}_j, \nln
\copro(\gen{F}_j)\eq \gen{F}_j\otimes q^{\gen{H}_j}+1\otimes \gen{F}_j.
\>
For the two fermionic generators $\gen{E}_2,\gen{F}_2$ an additional braiding with the 
generator $\gen{D}$ is introduced
\<
\label{eq:bradedcopr}
\copro(\gen{E}_2)\eq
\gen{E}_2\otimes 1+q^{-\gen{H}_2}q^{\gen{D}}\otimes \gen{E}_2,
\nln
\copro(\gen{F}_2)\eq
\gen{F}_2\otimes q^{\gen{H}_2}+q^{-\gen{D}}\otimes \gen{F}_2,
\nln
\copro(\gen{D})\eq\gen{D}\otimes 1+1\otimes\gen{D}.
\>
For convenience we have stated the coproduct of the central charge $\gen{D}$ 
which actually follows from the other coproducts.

\subsection{Fundamental Representation}

The above algebra has a family of four-dimensional fundamental representations.
Its vector space $\mathbb{V}$ has two bosonic and two fermionic directions. 
We assume it to be spanned by the four states
\[
\state{\phi^1},\state{\phi^2}\quad\mbox{and}\quad\state{\psi^1},\state{\psi^2}.
\]
The former two are bosonic and the latter two are fermionic.

\paragraph{Representation.}

The fundamental action of the Chevalley-Serre generators is given by%
\footnote{We have interchanged the states $\state{\psi^1}$ and $\state{\psi^2}$ 
as compared to \protect\cite{Beisert:2008tw}.}
\[\label{eq:fundquant}
\begin{array}{rclrclrclrcl}
\gen{H}_1\state{\phi^1}\eq -\state{\phi^1},&
\gen{H}_2\state{\phi^1}\eq -(C-\half)\state{\phi^1},&
\gen{E}_1\state{\phi^1}\eq q^{+1/2}\state{\phi^2},&
\gen{F}_2\state{\phi^1}\eq c\state{\psi^2}, \\[0.65ex]
\gen{H}_1\state{\phi^2}\eq +\state{\phi^2},&
\gen{H}_2\state{\phi^2}\eq -(C+\half)\state{\phi^2},&
\gen{E}_2\state{\phi^2}\eq a\state{\psi^1},&
\gen{F}_1\state{\phi^2}\eq q^{-1/2}\state{\phi^1}, \\[0.65ex]
\gen{H}_3\state{\psi^1}\eq +\state{\psi^1},&
\gen{H}_2\state{\psi^1}\eq -(C+\half)\state{\psi^1},&
\gen{E}_3\state{\psi^1}\eq q^{-1/2}\state{\psi^2},&
\gen{F}_2\state{\psi^1}\eq d\state{\phi^2}, \\[0.65ex]
\gen{H}_3\state{\psi^2}\eq -\state{\psi^2},&
\gen{H}_2\state{\psi^2}\eq -(C-\half)\state{\psi^2},&
\gen{E}_2\state{\psi^2}\eq b\state{\phi^1},&
\gen{F}_3\state{\psi^2}\eq q^{+1/2}\state{\psi^1}.
\end{array}
\]
The representation parameters $a,b,c,d$ 
must obey the constraint $(ad-qbc)(ad-q^{-1}bc)=1$.
They can be expressed in terms
of new parameters $\xpm{},\gamma$ as follows%
\footnote{As compared to \cite{Beisert:2008tw} 
we have rescaled $\gamma$ by $1/\sqrt{g}$ 
for later convenience.}
\[\label{eq:xpmparameters}
\begin{array}[b]{rclcrcl}
a\eq\gamma,
&\quad&
b\eq\displaystyle\frac{g\alpha}{\gamma}\,\frac{1}{\xm{}}\,\bigbrk{\xm{}-q^{2C-1}\xp{}},
\\[0.65ex]
c\eq\displaystyle \frac{i\gamma}{\alpha}\,\frac{q^{-C+1/2}}{\xp{}}\,,
&\quad&
d\eq\displaystyle \frac{ig}{\gamma}\,q^{C+1/2}\bigbrk{\xm{} -q^{-2C-1}\xp{}}.
\end{array}
\]
In terms of these parameters the constraint
implies the following quadratic relation between $\xpm{}$
\[\label{eq:xpmrel}
\frac{\xp{}}{q}+\frac{q}{\xp{}}-q\xm{}-\frac{1}{q\xm{}}
+ig(q-q^{-1})\lrbrk{\frac{\xp{}}{q\xm{}}-\frac{q\xm{}}{\xp{}}}=\frac{i}{g}\,.
\]
%

\paragraph{Central Charges.}

The central charge eigenvalues $D,C$ cannot be written unambiguously using $\xpm{}$,
but the combinations $q^{2D},q^{2C}$ are well-defined
\[\label{eq:q2c}
q^{2D}=\frac{\xp{}}{q\xm{}}\,,
\qquad
q^{2C}=q\,\frac{(q-q^{-1})/\xp{}-ig^{-1}}{(q-q^{-1})/\xm{}-ig^{-1}}
=q^{-1}\frac{(q-q^{-1})\xp{}+ig^{-1}}{(q-q^{-1})\xm{}+ig^{-1}}\,.
\]
The latter two expressions are equivalent upon \eqref{eq:xpmrel}.
Finally, the central charge eigenvalues $P,K$ follow 
from \eqref{eq:PKident}
\[\label{eq:PKU}
P=g\alpha\lrbrk{1-q^{2C}q^{2D}},
\qquad
K=
g\alpha^{-1}
\lrbrk{q^{-2C}-q^{-2D}}.
\]
The parameter $\gamma$ adjusts the normalisation
of bosons w.r.t.\ fermions in the representation; 
it is unphysical, but there is a preferable choice.

\paragraph{Fundamental R-Matrix.}

The quantum fundamental R-matrix $\rmat:
\mathbb{V}\otimes \mathbb{V}\to\mathbb{V}\otimes \mathbb{V}$ can be found by demanding
that it satisfies the cocommutativity relation 
\[
\coproop(\gen{J})\rmat=
\rmat\copro(\gen{J})
\]
for all generators $\gen{J}$ of the algebra,
where $\coproop$ is the opposite coproduct.
It turns out to be fully constrained by this 
relation up to one overall factor $R^0_{12}$.
The result is lengthy, and it can be found in \cite{Beisert:2008tw};
we refrain from reproducing it here.

\section{Classical Limit}
\label{sec:classical}

The classical limit of quantum-deformed R-matrices typically
consists in sending the deformation parameter $q$ to unity, $q\to 1$. 
For the undeformed R-matrix \cite{Beisert:2005tm}, however, the classical limit 
involves a large coupling constant, $g\to\infty$, 
cf.\ \cite{Klose:2006zd,Torrielli:2007mc}. 
Furthermore, the parameters for the fundamental representation
have to scale in a particular fashion such that $\xpm{}$ approach 
a common finite value, $\xpm{}\to x$.

We find that a reasonable classical limit consists in setting
\[\label{eq:qclass}
q=1+\frac{h}{2g}+\order{g^{-2}},
\]
with the inverse coupling constant $g^{-1}$ taking the role of the quantum parameter $\hbar$
\[\label{eq:gclass}
g\to\infty,
\]
while $h$ remains a finite deformation parameter even in the classical limit.
For later purposes we shall also introduce $h'$ as the combination 
\[
h'=\sqrt{1-h^2}\,.
\]
%

\subsection{Fundamental Representation}

For the parameters of the fundamental representation we 
assume the following classical limit%
\footnote{A simultaneous sign flip of $h'$ and $x$ changes nothing.}
\[\label{eq:xpmclass}
\xpm{}=(h'x-ih)\lrbrk{1\pm\frac{1}{2g}\,\frac{x(hx+ih')}{x^2-1}
+\order{g^{-2}}}.
\]
These obey the constraint \eqref{eq:xpmrel} up to the order given.
The coefficients $a,b,c,d$ in \eqref{eq:xpmparameters} 
then take the classical values
\[\label{eq:abcdclass}
a= \gamma,
\qquad
b= -\frac{\alpha (h'x-ih)(hx+ih')}{\gamma h'(x^2-1)}\,,
\qquad
c= \frac{i\gamma}{\alpha (h'x-ih)}\,,
\qquad
d= \frac{x(h'x-ih)}{\gamma h'(x^2-1)}\,.
\]
One can see that $ad-bc=1$ as desired for the classical limit $q\to1$.
The limit of the central charges $D,C,P,K$ in \eqref{eq:q2c,eq:PKU}
then follows as 
\[
D=-\half h^{-1}(z+1)q,\qquad
C=\half h^{-1}(z-1)q,\qquad
P=\alpha \conw q,\qquad
K=-\alpha^{-1} z \conw q,
\]
where $z$ and $q$ 
(the quantum parameter $q$ will not appear in the classical limit
and we can use the letter for a different purpose) are defined by 
\[\label{eq:zq}
z=\frac{ix}{(h'x-ih)(hx+ih')}\,,\qquad
q=\coni\frac{-(h'x-ih)(hx+ih')}{h'(x^2-1)}\,.
\]
%

\subsection{Fundamental r-Matrix}

We now take the classical limit 
\eqref{eq:qclass,eq:gclass,eq:xpmclass}
on the fundamental R-matrix $\rmat$ found in \cite{Beisert:2008tw}.
In the strict classical limit it reduces to the unity operator
and the first non-trivial order equals the 
fundamental classical r-matrix $r$
\[\label{eq:Rlimit}
(-A_{12}D_{12})^{-1/2}\rmat=1\otimes 1+\frac{h}{g}\,r+\order{g^{-2}}.
\]
Note that for definiteness we have multiplied the fundamental R-matrix by 
a combination of the coefficient functions $A_{12}$ and $D_{12}$ in \cite{Beisert:2008tw}. 
This removes the undetermined overall coefficient function $R^0_{12}$,
or it effectively fixes it to a convenient expression.


\begin{table}
\begin{eqnarray*}
r\state{\phi^1\phi^1}\eq
A_{12}\state{\phi^1\phi^1}
\nln
r\state{\phi^1\phi^2}\eq
\half (A_{12}+B_{12}+1)\state{\phi^2\phi^1}
+\half (A_{12}-B_{12})\state{\phi^1\phi^2}
-\half C_{12} \varepsilon_{\alpha\beta}\state{\psi^\alpha\psi^\beta}
\nln
r\state{\phi^2\phi^1}\eq
\half (A_{12}-B_{12})\state{\phi^2\phi^1}
+\half (A_{12}+B_{12}-1)\state{\phi^1\phi^2}
+\half C_{12}\varepsilon_{\alpha\beta}\state{\psi^\alpha\psi^\beta}
\nln
r\state{\phi^2\phi^2}\eq
A_{12}\state{\phi^2\phi^2}
\nn\\[10pt]
r\state{\psi^1\psi^1}\eq
-D_{12}\state{\psi^1\psi^1}
\nln
r\state{\psi^1\psi^2}\eq
-\half (D_{12}+E_{12}+1)\state{\psi^2\psi^1}
-\half (D_{12}-E_{12})\state{\psi^1\psi^2}
+\half F_{12}\varepsilon_{ab}\state{\phi^a\phi^b}
\nln
r\state{\psi^2\psi^1}\eq
-\half (D_{12}-E_{12})\state{\psi^2\psi^1}
-\half (D_{12}+E_{12}-1)\state{\psi^1\psi^2}
-\half F_{12}\varepsilon_{ab}\state{\phi^a\phi^b}
\nln
r\state{\psi^2\psi^2}\eq
-D_{12}\state{\psi^2\psi^2}
\nn\\[10pt]
r\state{\phi^a\psi^\beta}\eq
G_{12}\state{\phi^a\psi^\beta}
+H_{12}\state{\psi^\beta\phi^a}
\nln
r\state{\psi^\alpha\phi^b}\eq
K_{12}\state{\phi^b\psi^\alpha}
+L_{12}\state{\psi^\alpha\phi^b}
\nonumber
\end{eqnarray*}
\caption{The fundamental trigonometric r-matrix.}
\label{tab:rmatrix}
\end{table}
 
\begin{table}
\begin{eqnarray*}
A_{12}
=D_{12}
\eq
\frac{\quarter z_1+\quarter z_2+\quarter z_1 q_1 q_2^{-1}+\quarter z_2 q_2 q_1^{-1}}{z_1-z_2}
\nln
\half(A_{12}+B_{12}+1)
=\half(D_{12}+E_{12}+1)
\eq \frac{z_1}{z_1-z_2}
\nln
\half(A_{12}+B_{12}-1)
=\half(D_{12}+E_{12}-1)
\eq \frac{z_2}{z_1-z_2}
\nln
\half(A_{12}-B_{12})
=\half(D_{12}-E_{12})
\eq 
\frac{-\quarter z_1-\quarter z_2+\quarter z_1 q_1 q_2^{-1}+\quarter z_2 q_2 q_1^{-1}}{z_1-z_2}
\nln
C_{12}\eq
\frac{z_1a_1c_2-z_2a_2c_1}{z_1-z_2}
\nln
F_{12}\eq
\frac{z_1b_1d_2-z_2b_2d_1}{z_1-z_2}
\nln
G_{12}
=-L_{12}
\eq
\frac{-\quarter z_1 q_1 q_2^{-1}+\quarter z_2 q_2 q_1^{-1}}{z_1-z_2}
\nln
H_{12}\eq 
\frac{z_1a_1d_2-z_2b_2c_1}{z_1-z_2}
\nln
K_{12}\eq
\frac{-z_1b_1c_2+z_2a_2d_1}{z_1-z_2}
\nonumber
\end{eqnarray*}
\caption{The coefficients for the fundamental r-matrix.}
\label{tab:rcoeffs}
\end{table}

The resulting form of the fundamental classical r-matrix is given 
in \tabref{tab:rmatrix}.
It is determined by ten coefficient functions $A,B,C,D,E,F,G,H,K,L$.
Their values in the classical limit are given in \tabref{tab:rcoeffs}.
The matrix $r$ inherits the classical Yang--Baxter equation $\cybe{r}{r}=0$
from its quantum counterpart \cite{Beisert:2008tw}, where
\[\label{eq:cybeop}
\cybe{r}{s}:=
\comm{r_{12}}{s_{13}}
+\comm{r_{12}}{s_{23}}
+\comm{r_{13}}{s_{23}}.
\]

Taking a closer look at the coefficients we find four identities
among them: two linear ones 
\[\label{eq:rellin}
A-D=-B+E=G+L
\]
and two quadratic identities
\[\label{eq:relquad}
\quarter(A+B-1)(A+B+1)
=
\sfrac{1}{4}(3A-B)(3D-E)+4GL
=CF+HK.
\]
Note that we cannot in general claim that $A=D$
as suggested by \tabref{tab:rcoeffs}
because it follows only from our above choice
of prefactor $R^0_{12}$ in \eqref{eq:Rlimit}.
In other words, unlike the above four constraints
the latter one is not invariant under the shift
proportional to the identity matrix
\[\label{eq:identityshift}
\delta (A,B,C,D,E,F,G,H,K,L)\sim (+1,-1,0,-1,+1,0,+1,0,0,+1),
\]
which corresponds to changing the overall scattering phase. 
Altogether this reduces the $10$ coefficient functions 
to merely $6$ independent ones. This equals the number of free parameters:
$x_1$, $x_2$, $\gamma_1$, $\gamma_2$, $h$ and the freedom 
to shift by the identity matrix \eqref{eq:identityshift}.


\subsection{Lie Bialgebra}
\label{sec:loops}

In the following we shall derive a framework 
for the above r-matrix in terms of a Lie bialgebra.

\paragraph{Fundamental Representation and r-Matrix.}

First we would like to turn the fundamental r-matrix into
a universal one to make it applicable to arbitrary representations.
This is achieved by converting the operations in $r$, 
e.g.\ $\state{\phi^1\phi^2}\mapsto\state{\phi^2\phi^1}$,
to representations of symmetry generators, e.g.\ $-\gen{R}^{22}\otimes\gen{R}^{11}$,
acting individually on the two sites.

The operators $\gen{R},\gen{L},\gen{Q},\gen{S},\gencen,\gender$ 
are meant to mimic the fundamental representation of $\alg{gl}(2|2)$:
The two sets of $\alg{sl}(2)$ generators $\gen{R}^{ab}=\gen{R}^{ba}$ 
and $\gen{L}^{\alpha\beta}=\gen{L}^{\beta\alpha}$ 
act canonically on the two pairs of states $\state{\phi^a},\state{\psi^\alpha}$. 
The remaining operators are set up in analogy to \cite{Beisert:2007ty} 
to be able to reproduce the coefficients in \tabref{tab:rcoeffs}.
The action of the supercharges $\gen{Q}^{\alpha b}$ and $\gen{S}^{\alpha b}$ 
is specified through the parameters $a,b,c,d$. 
Finally, the action of the derivation $\gender$ 
and the central charge $\gencen$ involves $q$. 
Altogether the action reads
\[\label{eq:loopeval}
\begin{array}[b]{rclcrcl}
\gen{R}^{ab}\state{\phi^c}\eq \half\varepsilon^{bc}\state{\phi^a}+\half\varepsilon^{ac}\state{\phi^b},
&&
\gen{L}^{\alpha\beta}\state{\psi^\gamma}\eq \half\varepsilon^{\beta\gamma}\state{\psi^\alpha}-\half\varepsilon^{\alpha\gamma}\state{\psi^\beta},
\\[0.65ex]
\gen{Q}^{\alpha b}\state{\phi^c}\eq a\,\varepsilon^{bc}\state{\psi^\alpha},
&&
\gen{Q}^{\alpha b}\state{\psi^\gamma}\eq -b\,\varepsilon^{\alpha\gamma}\state{\phi^b},
\\[0.65ex]
\gen{S}^{\alpha b}\state{\phi^c}\eq -c\,\varepsilon^{bc}\state{\psi^\alpha},
&&
\gen{S}^{\alpha b}\state{\psi^\gamma}\eq d\,\varepsilon^{\alpha\gamma}\state{\phi^b},
\\[0.65ex]
\gencen\state{\phi^a}\eq \half q\,\state{\phi^a},
&&
\gencen\state{\psi^\alpha}\eq \half q\,\state{\psi^\alpha},
\\[0.65ex]
\gender\state{\phi^a}\eq -\half q^{-1}\state{\phi^a},
&&
\gender\state{\psi^\alpha}\eq +\half q^{-1}\state{\psi^\alpha}.
\end{array}
\]
The symbol $\varepsilon^{\cdot\cdot}$ 
is the antisymmetric $2\times 2$ matrix with $\varepsilon^{12}=+1$.
One can make contact with the fundamental 
representation of the quantum algebra in \eqref{eq:fundquant}
by means of the following identification 
with the Chevalley--Serre generators
\[
\begin{array}[b]{rclcrclcrcl}
\gen{H}_1\eq +2\gen{R}^{12} ,&&
\gen{E}_1\eq -\gen{R}^{22},&&
\gen{F}_1\eq +\gen{R}^{11},
\\[0.65ex]
\gen{H}_2\eq-h^{-1}(z-1)\gencen-\half\gen{H}_1-\half\gen{H}_3,&&
\gen{E}_2\eq+\gen{Q}^{11},&&
\gen{F}_2\eq-\gen{S}^{22},
\\[0.65ex]
\gen{H}_3\eq -2\gen{L}^{12} ,&&
\gen{E}_3\eq -\gen{L}^{22},&&
\gen{F}_3\eq +\gen{L}^{11}.
\end{array}\]
We are then led to the following form for the classical r-matrix 
from which the various coefficients in \tabref{tab:rcoeffs} are easily reproduced 
\<\label{eq:reval}
r\eq
+\frac{\half z_1+\half z_2}{z_1-z_2}\,
2\gen{R}^{12}\otimes \gen{R}^{12}
-\frac{z_1}{z_1-z_2}\,
\gen{R}^{22}\otimes \gen{R}^{11}
-\frac{z_2}{z_1-z_2}\,
\gen{R}^{11}\otimes \gen{R}^{22}
\nl
-\frac{\half z_1+\half z_2}{z_1-z_2}\,
2\gen{L}^{12}\otimes \gen{L}^{12}
+\frac{z_1}{z_1-z_2}\,
\gen{L}^{22}\otimes \gen{L}^{11}
+\frac{z_2}{z_1-z_2}\,
\gen{L}^{11}\otimes \gen{L}^{22}
\nl
-\frac{z_1}{z_1-z_2}\,
\varepsilon_{\alpha\gamma}\varepsilon_{bd}\gen{Q}^{\alpha b}\otimes \gen{S}^{\gamma d}
+\frac{z_2}{z_1-z_2}\,
\varepsilon_{\alpha\gamma}\varepsilon_{bd}\gen{S}^{\alpha b}\otimes \gen{Q}^{\gamma d}
\nl
-\frac{z_1}{z_1-z_2}\,
\gencen\otimes \gender
-\frac{z_2}{z_1-z_2}\,
\gender\otimes \gencen.
\>
%

\paragraph{Lie Brackets.}

Next we consider the commutators of the operators in \eqref{eq:loopeval}
to the end that they become the brackets of a Lie algebra 
and \eqref{eq:loopeval} define the fundamental representation.
From the way the indices are contracted in \eqref{eq:loopeval}, 
it is evident that $\gen{R}$ and $\gen{L}$ form two $\alg{sl}(2)$ algebras 
and that the generators $\gen{Q}$ and $\gen{S}$ 
transform in fundamental representations under these
\[\label{eq:RLeval}
\begin{array}[b]{rclcrcl}
\comm{\gen{R}^{ab}}{\gen{R}^{cd}}
\eq 
\varepsilon^{bc}\gen{R}^{ad}+\varepsilon^{ad}\gen{R}^{bc},
&\quad&
\comm{\gen{L}^{\alpha\beta}}{\gen{L}^{\gamma\delta}}
\eq 
\varepsilon^{\beta\gamma}\gen{L}^{\alpha\delta}+\varepsilon^{\alpha\delta}\gen{L}^{\beta\gamma},
\\[0.65ex]
\comm{\gen{R}^{ab}}{\gen{Q}^{\gamma d}}
\eq 
\half\varepsilon^{bd}\gen{Q}^{\gamma a}+\half\varepsilon^{ad}\gen{Q}^{\gamma b},
&\quad&
\comm{\gen{L}^{\alpha\beta}}{\gen{Q}^{\gamma d}}
\eq 
\half\varepsilon^{\beta\gamma}\gen{Q}^{\alpha d}+\half\varepsilon^{\alpha\gamma}\gen{Q}^{\beta d},
\\[0.65ex]
\comm{\gen{R}^{ab}}{\gen{S}^{\gamma d}}
\eq 
\half\varepsilon^{bd}\gen{S}^{\gamma a}+\half\varepsilon^{ad}\gen{S}^{\gamma b},
&\quad&
\comm{\gen{L}^{\alpha\beta}}{\gen{S}^{\gamma d}}
\eq 
\half\varepsilon^{\beta\gamma}\gen{S}^{\alpha d}+\half\varepsilon^{\alpha\gamma}\gen{S}^{\beta d}.
\end{array}
\]
The action of $\gen{Q},\gen{S},\gencen,\gender$ 
depends on the parameters $a,b,c,d,q$,
but their commutators 
can be written using only $z$ defined in \eqref{eq:zq}
\<\label{eq:DQeval}
\acomm{\gen{Q}^{\alpha b}}{\gen{Q}^{\gamma d}}
\eq  2\conw\alpha \varepsilon^{\alpha\gamma}\varepsilon^{bd} \gencen,
\nln
\acomm{\gen{Q}^{\alpha b}}{\gen{S}^{\gamma d}}
\eq 
-\varepsilon^{\alpha\gamma}\gen{R}^{bd}
+\varepsilon^{bd}\gen{L}^{\alpha\gamma}
-\varepsilon^{\alpha\gamma}\varepsilon^{bd}\conw h^{-1}(z-1) \gencen,
\nln
\acomm{\gen{S}^{\alpha b}}{\gen{S}^{\gamma d}}
\eq -2\conw \alpha^{-1}\varepsilon^{\alpha\gamma}\varepsilon^{bd}z\gencen,
\nln
\comm{\gender}{\gen{Q}^{\alpha b}}
\eq
\conw h^{-1}(z-1)\gen{Q}^{\alpha b}
+2\conw \alpha \gen{S}^{\alpha b},
\nln
\comm{\gender}{\gen{S}^{\alpha b}}
\eq
2\conw \alpha^{-1}z\gen{Q}^{\alpha b}
-\conw  h^{-1}(z-1)\gen{S}^{\alpha b}.
\>

Although the above generators and their relations are reminiscent of $\alg{gl}(2|2)$, 
they cannot form a Lie algebra as they stand.
The point is that the above commutators depend on the representation parameter $z$,
whereas the structure constants must be universal to the Lie algebra as a whole.
The way out is to consider instead the loop algebra of $\alg{gl}(2|2)[z,z^{-1}]$
in the way proposed in \cite{Beisert:2007ty}:
The variable $z$ can be interpreted as the formal loop variable
and the above action as an evaluation representation. 
Then the above commutation relations define Lie brackets
on the loop space $\alg{gl}(2|2)[z,z^{-1}]$.
The algebra is however not $\alg{gl}(2|2)[z,z^{-1}]$
because the above relations are not homogeneous in $z$. 
It is rather a non-trivial deformation of $\alg{gl}(2|2)[z,z^{-1}]$.

We observe that several of the coefficients appearing in 
\eqref{eq:DQeval} coincide. 
It turns out useful to combine these coefficients
as well as $a,b,c,d$ into $2\times 2$ matrices $W$ and $T$, respectively
\[\label{eq:matrices}
T=\matr{cc}{a&-b\\-c&d},
\qquad
W=\conw \matr{cc}{+h^{-1}(z-1)&2\alpha\\2\alpha^{-1} z&-h^{-1}(z-1)}.
\]
We note that $\det T=1$ and $\Tr W=0$.
Introducing a constant matrix $M$ we can write the relations required
to derive \eqref{eq:DQeval} in the compact form
\[\label{eq:Wdef}
TM=q WT,
\qquad
M=\matr{cc}{+1&0\\0&-1}.
\]
The above commutation relations \eqref{eq:DQeval} then read
\<\label{eq:DQmatrix}
\acomm{\gen{Q}^{\alpha b}}{\gen{Q}^{\gamma d}}
\eq \varepsilon^{\alpha\gamma}\varepsilon^{bd} W_{12}(z)\,\gencen,
\nln
\acomm{\gen{Q}^{\alpha b}}{\gen{S}^{\gamma d}}
\eq 
-\varepsilon^{\alpha\gamma}\gen{R}^{bd}
+\varepsilon^{bd}\gen{L}^{\alpha\gamma}
-\varepsilon^{\alpha\gamma}\varepsilon^{bd}W_{11}(z)\, \gencen,
\nln
\acomm{\gen{S}^{\alpha b}}{\gen{S}^{\gamma d}}
\eq -\varepsilon^{\alpha\gamma}\varepsilon^{bd} W_{21}(z)\,\gencen,
\nln
\comm{\gender}{\gen{Q}^{\alpha b}}
\eq
W_{11}(z)\,\gen{Q}^{\alpha b}
+W_{12}(z)\,\gen{S}^{\alpha b},
\nln
\comm{\gender}{\gen{S}^{\alpha b}}
\eq
W_{21}(z)\,\gen{Q}^{\alpha b}
+W_{22}(z)\,\gen{S}^{\alpha b},
\>
suggesting that $\gen{Q}^{\alpha b}$
and $\gen{S}^{\alpha b}$ 
form a two-component vector on which these matrices can act.
Note that the combinations for the brackets of supercharges
are naturally associated to the symmetric matrix
\[
-WN=\matr{cc}{+W_{12}&-W_{11}\\+W_{22}&-W_{21}}
\quad\mbox{with}\quad
N=\matr{cc}{0&+1\\-1&0}.
\]
%

\paragraph{Universal r-Matrix.}

The combinations of $z_1$ and $z_2$ appearing in \eqref{eq:reval}
are common for trigonometric classical r-matrices. 
We can split all of them into terms proportional to $z_1/(z_1-z_2)$ and $z_2/(z_1-z_2)$ 
\[\label{eq:rclassfunct}
r_{12}
=\frac{z_1}{z_1-z_2}\,s_{12}+\frac{z_2}{z_1-z_2}\,s_{21}
=s_{12}+\frac{z_2}{z_1-z_2}\,t_{12}.
\]
Here, $s$ and $t$ are following tensor products of generators
\<\label{eq:tr0}
s_{12}\eq
\gen{R}^{12}\otimes \gen{R}^{12}
-\gen{R}^{22}\otimes \gen{R}^{11}
-\gen{L}^{12}\otimes \gen{L}^{12}
+\gen{L}^{22}\otimes \gen{L}^{11}
-\varepsilon_{\alpha\gamma}\varepsilon_{bd}\gen{Q}^{\alpha b}\otimes \gen{S}^{\gamma d}
-\gencen\otimes \gender,
\nln
s_{21}\eq
\gen{R}^{12}\otimes \gen{R}^{12}
-\gen{R}^{11}\otimes \gen{R}^{22}
-\gen{L}^{12}\otimes \gen{L}^{12}
+\gen{L}^{11}\otimes \gen{L}^{22}
+\varepsilon_{\alpha\gamma}\varepsilon_{bd}\gen{S}^{\gamma d}\otimes\gen{Q}^{\alpha b}
-\gender\otimes \gencen,
\nln
t_{12}\eq
s_{12}+s_{21}
\nln\eq
-\varepsilon_{ac}\varepsilon_{bd}\gen{R}^{ab}\otimes \gen{R}^{cd}
+\varepsilon_{\alpha\gamma}\varepsilon_{\beta\delta}\gen{L}^{\alpha\beta}\otimes \gen{L}^{\gamma\delta}
\nl
-\varepsilon_{\alpha\gamma}\varepsilon_{bd}\gen{Q}^{\alpha b}\otimes \gen{S}^{\gamma d}
+\varepsilon_{\alpha\gamma}\varepsilon_{bd}\gen{S}^{\alpha b}\otimes \gen{Q}^{\gamma d}
-\gencen\otimes \gender
-\gender\otimes \gencen.
\>
The term $t$ is (graded) symmetric
and $r$ is (graded) anti-symmetric
\[\label{eq:rantisym}
t_{12}=t_{21}\,,\qquad
r_{12}+r_{21}=0.
\]

We can now consider the classical Yang--Baxter equation $\cybe{r}{r}=0$. 
The form of $r$ coincides with the conventional trigonometric r-matrix for 
the superalgebra $\alg{gl}(2|2)$ \cite{Leites:1984aa} 
for which the CYBE holds indeed.
Therefore the only violations could arise
from the deformations in \eqref{eq:DQmatrix}. 
We calculate the terms in $\cybe{r}{r}$ which consist of
one factor of $\gencen$ and two supercharges.
These turn out to vanish if the following
three equations hold for 
$F_1(z)=zW_{12}(z)$, $F_2(z)=W_{11}(z)$ and $F_3(z)=z^{-1}W_{21}(z)$
\[
 \frac{F_{k}(z_1)}{(z_1-z_2)(z_1-z_3)}
-\frac{F_{k}(z_2)}{(z_1-z_2)(z_2-z_3)}
+\frac{F_{k}(z_3)}{(z_1-z_3)(z_2-z_3)}
=0\,.
\]
Setting $z_3=0$ the equation reduces to 
\[
\frac{F_k(z_1)-F_k(0)}{z_1}=\frac{F_k(z_2)-F_k(0)}{z_2}\,.
\]
This implies that $F_k(z)$ must be a polynomial of degree 1
which is indeed a solution of the above equation
and which is also true for all matrix elements in \eqref{eq:matrices}.
Therefore the CYBE is fulfilled, and the r-matrix 
enhances the loop algebra to a triangular Lie bialgebra.
Note that the above three conditions also guarantee that 
the algebra has a positive, a negative and a Cartan subalgebra,
see \eqref{eq:decompose} for more details.

\subsection{Loop Level Form}
\label{sec:levels}

In order to define the loop algebra more rigorously, 
we shall provide an alternative presentation in terms of 
the generators 
at definite levels of the loop algebra 
\[
\gen{J}_n\simeq z^n \gen{J}.
\] 
This description of the loop algebra is instructive,
and it has in fact a slightly different bialgebra structure.
Nevertheless in the remainder of the paper 
we shall mostly employ the functional description introduced above.

\paragraph{Lie Brackets.}

Based on the above operators 
$\gen{J}\in\spn{\gen{R},\gen{L},\gen{Q},\gen{S},\gencen,\gender}=\alg{gl}(2|2)$
we define an algebra spanned by $\gen{J}_n$ for $n\in\Integers$.
The vector space of the algebra is the one of $\alg{gl}(2|2)[z,z^{-1}]$,
but the Lie brackets are deformed:
The brackets involving the two sets of $\alg{sl}(2)$ generators $\gen{R}$ and $\gen{L}$
are precisely as in $\alg{gl}(2|2)[z,z^{-1}]$
\[\label{eq:braRL}
\begin{array}[b]{rclcrcl}
\bigcomm{\gen{R}^{ab}_m}{\gen{R}^{cd}_n}
\eq 
\varepsilon^{bc}\gen{R}^{ad}_{m+n}+\varepsilon^{ad}\gen{R}^{bc}_{m+n},
&\quad&
\bigcomm{\gen{L}^{\alpha\beta}_m}{\gen{L}^{\gamma\delta}_n}
\eq 
\varepsilon^{\beta\gamma}\gen{L}^{\alpha\delta}_{m+n}+\varepsilon^{\alpha\delta}\gen{L}^{\beta\gamma}_{m+n},
\\[0.65ex]
\bigcomm{\gen{R}^{ab}_m}{\gen{Q}^{\gamma d}_n}
\eq 
\half\varepsilon^{bd}\gen{Q}^{\gamma a}_{m+n}+\half\varepsilon^{ad}\gen{Q}^{\gamma b}_{m+n},
&\quad&
\bigcomm{\gen{L}^{\alpha\beta}_m}{\gen{Q}^{\gamma d}_n}
\eq 
\half\varepsilon^{\beta\gamma}\gen{Q}^{\alpha d}_{m+n}+\half\varepsilon^{\alpha\gamma}\gen{Q}^{\beta d}_{m+n},
\\[0.65ex]
\bigcomm{\gen{R}^{ab}_m}{\gen{S}^{\gamma d}_n}
\eq 
\half\varepsilon^{bd}\gen{S}^{\gamma a}_{m+n}+\half\varepsilon^{ad}\gen{S}^{\gamma b}_{m+n},
&\quad&
\bigcomm{\gen{L}^{\alpha\beta}_m}{\gen{S}^{\gamma d}_n}
\eq 
\half\varepsilon^{\beta\gamma}\gen{S}^{\alpha d}_{m+n}+\half\varepsilon^{\alpha\gamma}\gen{S}^{\beta d}_{m+n}.
\end{array}
\]
Only the brackets between supercharges $\gen{Q}$, $\gen{S}$
and the derivation $\gender$ are modified. 
They follow from the above commutators 
for the fundamental representation
\eqref{eq:DQeval}
where the variable $z$ is interpreted as a shift by one level
\<\label{eq:braDQ}
\bigacomm{\gen{Q}^{\alpha b}_m}{\gen{Q}^{\gamma d}_n}
\eq  2\conw\alpha \varepsilon^{\alpha\gamma}\varepsilon^{bd} \gencen_{m+n},
\nln
\bigacomm{\gen{Q}^{\alpha b}_m}{\gen{S}^{\gamma d}_n}
\eq 
-\varepsilon^{\alpha\gamma}\gen{R}^{bd}_{m+n}
+\varepsilon^{bd}\gen{L}^{\alpha\gamma}_{m+n}
-\varepsilon^{\alpha\gamma}\varepsilon^{bd}\conw h^{-1}\bigbrk{\gencen_{m+n+1}-\gencen_{m+n}},
\nln
\bigacomm{\gen{S}^{\alpha b}_m}{\gen{S}^{\gamma d}_n}
\eq -2\conw \alpha^{-1} \varepsilon^{\alpha\gamma}\varepsilon^{bd}\gencen_{m+n+1},
\nln
\bigcomm{\gender_m}{\gen{Q}^{\alpha b}_n}
\eq
\conw h^{-1}\bigbrk{\gen{Q}^{\alpha b}_{m+n+1}-\gen{Q}^{\alpha b}_{m+n}}
+2\conw \alpha \gen{S}^{\alpha b}_{m+n},
\nln
\bigcomm{\gender_m}{\gen{S}^{\alpha b}_n}
\eq
2\conw \alpha^{-1}\gen{Q}^{\alpha b}_{m+n+1}
-\conw  h^{-1}\bigbrk{\gen{S}^{\alpha b}_{m+n+1}-\gen{S}^{\alpha b}_{m+n}}.
\>
The remaining unspecified Lie brackets are trivial.
Altogether the Jacobi identities are satisfied as
can be confirmed explicitly.
The algebra has a family of four-dimensional evaluation representations 
with $\gen{J}_n\simeq z^n \gen{J}$ 
and the action of $\gen{J}$ specified in \eqref{eq:loopeval}.


\paragraph{Universal r-Matrix.}

The functional r-matrix in \eqref{eq:rclassfunct} 
can be cast into the loop level form.
To that end one expands the above function of $z$'s into a geometric series%
\footnote{This formula represents an analytic continuation
of the series, see below for additional distributional contributions.}
\[
\frac{z_2}{z_1-z_2}=
\sum_{k=1}^{\infty} \lrbrk{\frac{z_2}{z_1}}^{k}\,.
\]
The resulting r-matrix then reads explicitly
\<\label{eq:rclass}
r\eq 
\gen{R}^{12}_0\otimes \gen{R}^{12}_0
-\gen{R}^{22}_0\otimes \gen{R}^{11}_0
-\gen{L}^{12}_0\otimes \gen{L}^{12}_0
+\gen{L}^{22}_0\otimes \gen{L}^{11}_0
-\varepsilon_{\alpha\gamma}\varepsilon_{bd}\gen{Q}^{\alpha b}_0\otimes \gen{S}^{\gamma d}_0
-\gencen_0\otimes \gender_0
\nl
+\sum_{k=1}^\infty \Big[
-\varepsilon_{ac}\varepsilon_{bd}\gen{R}^{ab}_{-k}\otimes \gen{R}^{cd}_{+k}
+\varepsilon_{\alpha\gamma}\varepsilon_{\beta\delta}\gen{L}^{\alpha\beta}_{-k}\otimes \gen{L}^{\gamma\delta}_{+k}
\nl\qquad\qquad
-\varepsilon_{\alpha\gamma}\varepsilon_{bd}\gen{Q}^{\alpha b}_{-k}\otimes \gen{S}^{\gamma d}_{+k}
+\varepsilon_{\alpha\gamma}\varepsilon_{bd}\gen{S}^{\alpha b}_{-k}\otimes \gen{Q}^{\gamma d}_{+k}
\nl\qquad\qquad
-\gencen_{-k}\otimes \gender_{+k}
-\gender_{-k}\otimes \gencen_{+k}
\Big].
\>

This r-matrix defines a quasi-triangular Lie bialgebra:
First of all the symmetric part of $r$ 
equals
\[\label{eq:rsyminfsum}
r_{12}+r_{21}=\sum_{k=-\infty}^\infty \lrbrk{\frac{z_2}{z_1}}^k t,
\]
which is an invertible quadratic invariant of the algebra.
It is straight-forward to convince oneself of this fact.
Note that this is slightly different than 
in the functional form where $r_{12}+r_{21}=0$, cf.\ \eqref{eq:rantisym}.
Consequently, the resulting algebra is merely \emph{quasi}-triangular.

Furthermore, the classical Yang--Baxter equation $\cybe{r}{r}=0$, 
see \eqref{eq:cybeop}, holds.
This is not as easily seen, 
in particular it does not follow right away from 
the discussion at the end of \secref{sec:loops}
because the expansion into loop levels introduces
slight modifications, cf.\ \eqref{eq:rantisym}
vs.\ \eqref{eq:rsyminfsum}.
The CYBE eventually follows 
from the triangular decomposition of $\alg{gl}(2|2)[z,z^{-1}]$ 
into a positive, negative and Cartan subalgebra 
$\alg{g}^+\oplus \alg{g}^- \oplus \alg{g}^0$
with 
\<\label{eq:decompose}
\alg{g}^+\eq 
\spn{\gen{R}^{11}_0,\gen{L}^{11}_0,\gen{S}_0,\gender_0}
\oplus z\,\alg{gl}(2|2)[z],
\nln
\alg{g}^0\eq 
\spn{\gen{R}^{12}_0,\gen{L}^{12}_0},
\nln
\alg{g}^-\eq 
\spn{\gen{R}^{22}_0,\gen{L}^{22}_0,\gen{Q}_0,\gencen_0}
\oplus z^{-1}\,\alg{gl}(2|2)[z^{-1}].
\>
The crucial observation which ensures quasi-triangularity 
is that the r-matrix \eqref{eq:rclass} 
belongs to the following subspace
(more precisely its compactification)
\[
r\in (\alg{g}^-\otimes \alg{g}^+) \oplus (\alg{g}^0\otimes \alg{g}^0).
\]
This r-matrix takes the form of the classical double of $\alg{g}^+\oplus\alg{g}^0$
divided by the centre generated by a combination of $\alg{g}^0$ and its dual.
Alternatively, one can say that 
the decomposition $(\alg{g}^+\oplus \alg{g}^0\oplus\alg{g}^-,\alg{g}^+\oplus \alg{g}^0,\alg{g}^-\oplus \alg{g}^0)$ 
is a Manin triple up to the double appearance of the 
Cartan subalgebra $\alg{g}^0$. 

\paragraph{Distributions on the Complex Plane.}

To convert between the above two pictures for loop algebras
one conventionally uses the geometric series 
\[\label{eq:geosum}
g(z):=\sum_{n=0}^\infty z^n,
\qquad
g(z)=\frac{1}{1-z}\mbox{ for }|z|<1.
\]
It is convenient to continue the function $g(z)$ analytically 
to all $z\neq 1$, but some care is required because 
it actually introduces inconsistencies:
Consider the contour integral of $z^k g(z)$
for a circle of radius $r$ around the origin.
One would like to obtain the following result 
for the geometric series 
(i.e.\ when performing the integral prior to the infinite sum)
\[
\frac{1}{2\pi i}\,\oint_r z^k g(z)\, dz=\delta_{k<0}.
\]
When analytically continuing the series $g(z)$ to all $z\neq 1$ one obtains a 
different result
\[
\frac{1}{2\pi i}\,\oint_r \frac{z^k\,dz}{1-z}=
\begin{cases}
+\delta_{k<0}&\mbox{for }r<1,\\
-\delta_{k\geq 0}&\mbox{for }r>1.
\end{cases}\]
The difference between the two integrals equals $-1$ for $r>1$ irrespectively of the value of $k$. 
Such a term can be thought of as to originate from a distributional term $\delta_{a,b}(z)$
which is supported on a curve between $a$ and $b$.%
\footnote{The distributions become somewhat more familiar,
see e.g.\ \cite{Faddeev:1987ph},
when the variables and integrations are restricted to the unit circle, $|z|=1$.}
The distribution is defined such that 
for each (directed) crossing of the contour through the supporting curve,
the distribution contributes the value of the integrand at $z=0$.
Now the distributional result of the geometric series reads
\[
g(z)=\frac{1}{1-z}+2\pi i\delta_{0,\infty}(z-1),
\]
and the extra term w.r.t.\ \eqref{eq:geosum} is what
reduces a triangular algebra to a \emph{quasi}-triangular one.
Now $g(z)$ has a cut on the positive real axis extending from $z=1$ to $z=\infty$.
Each crossing of the cut from the lower towards the upper half plane 
contributes the value of the integrand at $z=1$
\[
\frac{1}{2\pi i}\,\oint_r z^k\, 2\pi i\delta_{0,\infty}(z-1)\,dz=
\begin{cases}
0&\mbox{for }r<1,\\
1&\mbox{for }r>1.
\end{cases}\]

The quadratic invariant requires a geometric series
over both positive and negative powers.
For such series \eqref{eq:rsyminfsum}
the analytic contribution vanishes exactly, 
while a distributional contribution remains
\[
\sum_{n=-\infty}^\infty z^n=g(z)+g(1/z)-1=
2\pi i\delta_{0,\infty}(z-1)
+2\pi i\delta_{0,\infty}(1/z-1)
=
2\pi i\delta_{-1,\infty}(z-1).
\]
We have made use of proper transformation rules for this distribution
which are analogous to those for delta functions. 
Here the resulting branch cut extends from $z=0$ to $z=\infty$,
and for each crossing it contributes the value of the integrand at $z=1$.
In the remainder of the paper we will only make reference to this 
type of distribution, written in the form
\[\label{eq:doublegeometric}
\sum_{n=-\infty}^\infty \lrbrk{\frac{z_1}{z_2}}^n=
2\pi i\delta_{-1,\infty}(z_1/z_2-1)
=:2\pi iz_1\delta(z_1-z_2).
\]
The latter is a convenient abbreviation of the former distribution:
Here the cut extends from $z_1=0$ to $z_1=\infty$
or alternatively from $z_2=\infty$ to $z_2=0$.

\section{Affine Extension}
\label{sec:affine}

A loop algebra can be extended by 
one derivation $\genaffder$ and one central charge $\genaffcen$ 
to an affine (Kac--Moody) Lie algebra.
Here we show that our deformed loop algebra also admits such
an affine extension.

\subsection{Example}

We shall use the example of $\alg{sl}(2)$ to illustrate the construction
of the affine extension.
The derivation $\genaffder$ is defined as the following derivative w.r.t.\ $z$ 
\[\label{eq:derdiff}
\genaffder=\frac{zd}{dz}\,.
\]
Put differently, $\genaffder$ generates a scaling transformation of $z$.
Alternatively we can define $\genaffder$ through its action on 
the loop variable $z$ and the base generators $\gen{R}^{ab}$ 
\[\comm{\genaffder}{z}=z,
\qquad
\comm{\genaffder}{\gen{R}^{ab}}=0.
\]
The central charge appears in the brackets as follows
\[
\bigcomm{f(z)\gen{R}^{ab}}{g(z)\gen{R}^{cd}}
=
f(z)g(z)\bigbrk{\varepsilon^{bc}\gen{R}^{ad}+\varepsilon^{ad}\gen{R}^{bc}}
-\half\lrbrk{\varepsilon^{ac}\varepsilon^{bd}+\varepsilon^{ad}\varepsilon^{bc}}
\frac{1}{2\pi i}\oint \bigbrk{f(z)dg(z)}
\,\genaffcen.
\]
Here the contour integral winds once around $z=0$ (or $z=\infty$).
Finally, we can write the quadratic invariant 
using the delta distribution in \eqref{eq:doublegeometric}
\[
\hat t=
-2\pi iz_1\delta(z_1-z_2) 
\varepsilon_{ac}\varepsilon_{bd}\gen{R}^{ab}\otimes \gen{R}^{cd}
-\genaffder\otimes\genaffcen-\genaffcen\otimes\genaffder.
\]

The above construction can be generalised 
straight-forwardly to any loop algebra,
but for our deformed loop algebra some more work is needed
because of the non-homogeneous structure of the loop levels in \eqref{eq:DQeval}.

\subsection{Derivation}

Now we have to generalise the brackets with the affine derivation to 
all generators of our loop algebra. 
First of all, it acts on the loop parameter $z$ 
as a scaling transformation
\[\label{eq:affderloops}
\comm{\genaffder}{z}=z
\quad
\mbox{or}
\quad
\genaffder\simeq \frac{zd}{dz}\,.
\]
The derivations of the two sets of $\alg{sl}(2)$ 
generators $\gen{R}$ and $\gen{L}$ take the standard form
\[\label{eq:RLaffder}
\comm{\genaffder}{\gen{R}^{ab}}=0,\qquad 
\comm{\genaffder}{\gen{L}^{\alpha\beta}}=0.
\]
For the remaining generators $\gen{Q},\gen{S},\gencen,\gender$ 
we can gain inspiration from the fundamental representation in \eqref{eq:loopeval}.
As compared to the fundamental representation of 
the undeformed $\alg{gl}(2|2)$, 
the representations of $\gen{Q}$ and $\gen{S}$ (as a 2-vector) 
are rotated by the $\grp{SL}(2)$ matrix $T$ in \eqref{eq:matrices}
\cite{Hofman:2006xt}.
Furthermore the action of $\gencen$ and $\gender$ is scaled by $q$
w.r.t.\ the undeformed $\alg{gl}(2|2)$.
The parameters $a,b,c,d,q$ depend on $x$ which is related to $z$ via \eqref{eq:zq}. 
The derivation $\genaffder$ transforms $z$ according to \eqref{eq:derdiff},
hence it modifies the matrix $T$. 
The brackets of the derivation with the supercharges 
must reflect this transformation in order to find a suitable
representation of $\genaffder$. 
We are thus led to the following combinations
\[\label{eq:UVdef}
z\,\frac{dT}{dz}\,T^{-1}
=U+f(z)W,
\qquad
\frac{z}{q}\,\frac{dq}{dz}=V,
\]
with%
\footnote{The conversion to levels of the loop algebra
along the lines of \secref{sec:levels}
is somewhat problematic due to the presence
of poles in $U(z)$ and $V(z)$ at $z\neq0,\infty$.
This issue deserves further investigations.}
\[\label{eq:UVmatrix}
U=
\frac{1}{z+z^{-1}-2+4h^2}\matr{cc}
{-h^2&+h\alpha\\-h\alpha^{-1}&+h^2},
\qquad
V=
-\frac{z-1+2h^2}{z+z^{-1}-2+4h^2}\,.
\]
The precise functional form of $\gamma$ 
influences the undetermined function $f(z)$. 
For $f(z)=0$ we get a reasonably simple final expression 
corresponding to the choice
\[
\gamma\simeq \frac{h'x-ih}{h'\sqrt{x^2-1}}\,.
\]
The matrix $U$ now appears as the derivation 
of the two-vector of the bare supercharges $\gen{Q}$ and $\gen{S}$. 
Altogether the derivations are specified by
\<\label{eq:DQaffder}
\comm{\genaffder}{\gen{Q}^{\alpha b}}
\eq
U_{11}(z)\,\gen{Q}^{\alpha b}
+U_{12}(z)\,\gen{S}^{\alpha b},
\nln
\comm{\genaffder}{\gen{S}^{\alpha b}}
\eq
U_{21}(z)\,\gen{Q}^{\alpha b}
+U_{22}(z)\,\gen{S}^{\alpha b},
\nln
\comm{\genaffder}{\gencen}
\eq +V(z)\,\gencen,
\nln
\comm{\genaffder}{\gender}
\eq -V(z)\,\gender.
\>
Note that the ambiguity in \eqref{eq:UVdef} 
corresponds to shifting $\genaffder$ by $f(z)\gender$,
cf.\ \eqref{eq:DQmatrix};
nothing is lost by making a specific choice as the above.
The Jacobi identities require
\[
z\,\frac{dW}{dz}=\comm{U}{W}-VW,
\]
which follows by combining \eqref{eq:Wdef} with \eqref{eq:UVdef}.
As an aside, we note that the derivation $\genaffder$ can be extended to 
a Virasoro algebra $\genaffder_n=z^n\genaffder$
with a new central charge $\gen{c}$, 
but we will not make use of it here.

\subsection{Central Charge}

The central charge appears in the brackets of the two sets of $\alg{sl}(2)$ 
generators in the standard fashion
\<\label{eq:RLcentral}
\bigcomm{f(z)\gen{R}^{ab}}{g(z)\gen{R}^{cd}}
\eq
f(z)g(z)\bigbrk{\varepsilon^{bc}\gen{R}^{ad}-\varepsilon^{ad}\gen{R}^{bc}}
\nl
-\half\lrbrk{\varepsilon^{ac}\varepsilon^{bd}+\varepsilon^{ad}\varepsilon^{bc}}
\frac{1}{2\pi i}\oint \bigbrk{f(z)dg(z)}\,\genaffcen,
\nln
\bigcomm{f(z)\gen{L}^{\alpha\beta}}{g(z)\gen{L}^{\gamma\delta}}
\eq
f(z)g(z)\bigbrk{\varepsilon^{\beta\gamma}\gen{L}^{\alpha\delta}-\varepsilon^{\alpha\delta}\gen{L}^{\beta\gamma}}
\nl
+\half\lrbrk{\varepsilon^{\alpha\gamma}\varepsilon^{\beta\delta}+\varepsilon^{\alpha\delta}\varepsilon^{\beta\gamma}}
\frac{1}{2\pi i}\oint \bigbrk{f(z)dg(z)}\,\genaffcen.
\>
For the remaining generators $\gen{Q},\gen{S},\gencen,\gender$ 
the brackets leading to the central charge have to be adjusted 
to the deformations in \eqref{eq:DQaffder}.
There are several ways to derive a central charge 
for the above loop algebra. A very convenient method
consists in demanding invariance of the 
quadratic invariant, cf.\ \eqref{eq:doublegeometric},
\[
\hat t=
2\pi i z_1\delta(z_1-z_2)t
-\genaffcen\otimes\genaffder
-\genaffder\otimes\genaffcen
\]
where $t$ is given in \eqref{eq:tr0}.
The invariance under the loop generators
requires a balancing of two types of terms:
The contributions from brackets with $\genaffder$ 
must cancel the contribution from brackets 
proportional to the central charge. 
One can easily figure out the central charge
contributions complementary to \eqref{eq:DQaffder}
%
\<\label{eq:DQcentral}
\bigacomm{f(z)\gen{Q}^{\alpha b}}{g(z)\gen{Q}^{\gamma d}}
\eq 
\varepsilon^{\alpha\gamma}\varepsilon^{bd} f(z)g(z)W_{12}(z)\, \gencen
\nl
+\varepsilon^{\alpha\gamma}\varepsilon^{bd}
\frac{1}{2\pi i}\oint \lrbrk{f(z)g(z)U_{12}(z)\,\frac{dz}{z}}\,\genaffcen,
\nln
\bigacomm{f(z)\gen{Q}^{\alpha b}}{g(z)\gen{S}^{\gamma d}}
\eq
f(z)g(z)
\lrbrk{ 
-\varepsilon^{\alpha\gamma}\gen{R}^{bd}
+\varepsilon^{bd}\gen{L}^{\alpha\gamma}
-\varepsilon^{\alpha\gamma}\varepsilon^{bd}W_{11}(z)\, \gencen
}
\nl
+\varepsilon^{\alpha\gamma}\varepsilon^{bd}
\frac{1}{2\pi i}\oint 
\lrbrk{f(z)dg(z)-f(z)g(z)U_{11}(z)\,\frac{dz}{z}}
\,\genaffcen,
\nln
\bigacomm{f(z)\gen{S}^{\alpha b}}{g(z)\gen{S}^{\gamma d}}
\eq 
-\varepsilon^{\alpha\gamma}\varepsilon^{bd}f(z)g(z) W_{21}(z)\, \gencen
\nl
+\varepsilon^{\alpha\gamma}\varepsilon^{bd}
\frac{1}{2\pi i}\oint 
\lrbrk{-f(z)g(z)U_{21}(z)\,\frac{dz}{z}}
\,\genaffcen,
\nln
\bigcomm{f(z)\gencen}{g(z)\gender}
\eq 
-\frac{1}{2\pi i}\oint 
\lrbrk{f(z)dg(z)-V(z) f(z)g(z)\,\frac{dz}{z}}
\,\genaffcen.
\>
Note that the value of the above integrals depends on the
choice of contours. Conventionally one assumes that
the functions $f(z)$ and $g(z)$ are holomorphic
except at $z=0$ and $z=\infty$. In that case the contour can take any path 
that winds once around $z=0$ and $z=\infty$.
Here the functions $W(z)$ and $V(z)$ in \eqref{eq:UVmatrix} 
introduce two extra poles $z^*_\pm$. 
These could be used to define two additional central charges.
It appears that they behave much like $\delta(z-z^*_\pm)\gencen$
and therefore there may be no need to enlarge the algebra further.
The issues of how to put the contours and how to define the 
affine central charge(s) need further investigations.

\subsection{Affine r-Matrix}

The affine extension of the r-matrix in \eqref{eq:rclassfunct} reads
\[
\hat r=r-\genaffcen\otimes\genaffder.
\]
In the presence of $\genaffcen$ the additional term is needed 
to fulfil the CYBE.
Note that one is free to add an antisymmetric term proportional to 
$\genaffcen\otimes\genaffder-\genaffder\otimes\genaffcen$ 
to the above $\hat{r}$
\cite{Reshetikhin:1990ep}.

A curious feature of the r-matrix is that it is not 
invariant under the affine derivation $\genaffder$. 
This is because the coefficients $U,V$ in 
\eqref{eq:UVmatrix} of the action \eqref{eq:DQaffder}
depend on $z$. Effectively this implies that the 
r-matrix is not a function of $z_2/z_1$ alone, but it 
depends separately on $z_1$ and $z_2$. 
Correspondingly, the cobracket of $\genaffder$ becomes non-trivial. 
 
A possible benefit of the affine extension is that it may 
add further constraints on the r-matrix. 
Without the extension it is possible to add to $r$ terms of
the form
\[
z_1^m z_2^n\gencen\otimes\gencen
\]
because $\gencen$ is a central element and hence it cannot be 
seen within the CYBE. 
One could also view the deformation 
as a deformation of $z^m\gender$ by $z^n\gencen$ 
which affects only the coalgebra but not the algebra.
In the presence of the affine extensions such deformations 
may no longer be possible because $\gencen$ is no longer in the centre;
it has non-trivial brackets with $\genaffder$ and $\gender$.
Thus it would be interesting to derive constraints
on the permissible deformations of $r$ by $\gencen\otimes\gencen$.

\subsection{Fundamental Representation}

Let us reconsider the fundamental evaluation representation 
\eqref{eq:loopeval}.
The regular loop generators act on a four-dimensional space
spanned by $\state{\phi^a}$ and $\state{\psi^\alpha}$. 
Conversely, the derivation $\genaffder$ 
acts as a scaling transformation \eqref{eq:derdiff} for the parameter $z$
which is related to the representation parameter $x$ through \eqref{eq:zq}.
Consequently we must promote the states to fields 
$\state{\phi^a,x}$ and $\state{\psi^\alpha,x}$ so that $\genaffder$ can act on them. 
The representation of the affine algebra is therefore infinite-dimensional,
and it naturally models a field on a one-dimensional mass shell. 
Effectively, the affine derivation corresponds to a (Lorentz) boost
of the mass shell. 
In this case the cobrackets for $\genaffder$ are non-trivial 
and therefore Lorentz symmetry must be considered as deformed.

In the picture of fields $\state{\phi^a,x}$ and $\state{\psi^\alpha,x}$, 
one can get a clearer understanding of the role of the generator $z^k\gen{B}$
(note that $x$ is related to $z$): 
Eq.\ \eqref{eq:loopeval} suggests that it induces a $x$-dependent (i.e.\ gauge) 
transformation for the normalisation of bosons w.r.t.\ fermions. 
The role of $\gamma$ (which can now depend on $x$) is related: 
It serves as a (functional) parameter of the representation, 
and it fixes a particular normalisation for it.

This evaluation-type representation clearly has vanishing central charge $\genaffcen\simeq 0$.
However, there surely exist representations with non-vanishing central charge,
such as highest-weight representations.
In the physical context these may correspond to vertex operators.
It would be interesting to investigate charged representations of this algebra.

\section{Discrete Symmetries}
\label{sec:discrete}

Before we continue with particular limiting cases of the classical r-matrix, 
we shall discuss some of its discrete symmetries.
These will help us understand the limits better 
and also relate some cases to others. 

\subsection{Conjugation}
\label{sec:conjugation}

The map \eqref{eq:zq} between $x$ and $z$ is quadratic and thus 2:1. 
The underlying reason for this property is that for unitary 
superalgebras there are four conjugate fundamental representations.
In the present algebra, however, there is just a one-parameter family of 
fundamental representations parametrised through $x$.
For each value of $z$ there are two values of $x$ corresponding to 
a pair of representations and its conjugate:
That this is possible in the first place is a special property of $\alg{sl}(2|2)$.
The representation of each $\alg{sl}(2)$ subalgebra is fundamental;
as such it is self-conjugate under transposition and
conjugation by an antisymmetric $2\times 2$ matrix $\varepsilon$,
i.e.\ for a traceless $2\times 2$ matrix $E$ one has
\[
E'=-\varepsilon E^\trans \varepsilon^{-1}=E.
\]
However, the representation of the remaining generators 
is not self-conjugate under the combined map for a $4\times 4$ 
supermatrix $E$ written in $2\times2$ blocks
\[\label{eq:stransconj}
E'=-\matr{c|c}{\varepsilon&0\\\hline0&\varepsilon}
E^\strans
\matr{c|c}{\varepsilon^{-1}&0\\\hline0&\varepsilon^{-1}}.
\]
The representation $E$ is parametrised through $x$ and $\gamma$. 
The two values of $x$ corresponding to conjugate representations
are related by inversion \cite{Beisert:2008tw}
\[
x'=\frac{1}{x}\,,
\qquad
z'=z.
\]
The parameters \eqref{eq:abcdclass,eq:zq,eq:matrices} 
for the fundamental representation map according to
\[
T'=T\matr{cc}{0&-1\\+1&0},\qquad
q'=-q,\qquad
\gamma'=\frac{\alpha}{\gamma}\,\frac{(h'x-ih)(hx+ih')}{h'(x^2-1)}\,.
\]
Note that the matrix multiplying $T$ 
corresponds to the supertranspose operation
which is $\Integers_4$ periodic.
So for each value of $x$ there are two representations 
which differ in sign for the odd generators
corresponding to a total of four fundamentals in superalgebras.
See also \secref{sec:uniform} for further comments.

This transformation involves only representations 
and thus it can be applied to each of the two sites
of the fundamental r-matrix individually.
Under such a crossing transformation of $x_1,\gamma_1$ the coefficients in 
\tabref{tab:rcoeffs} permute as follows
\[
\begin{array}[b]{rclcrcl}
A_{\bar 12}\eq -\half (A_{12}-B_{12}),
&\quad&
D_{\bar 12}\eq-\half (D_{12}-E_{12}),
\\[0.65ex]
\half(A_{\bar 12}-B_{\bar 12})\eq -A_{12},
&\quad&
\half(D_{\bar 12}-E_{\bar 12})\eq-D_{12},
\\[0.65ex]
G_{\bar 12}\eq L_{12},
&\quad&
L_{\bar 12}\eq G_{12},
\\[0.65ex]
H_{\bar 12}\eq -F_{12},
&\quad&
F_{\bar 12}\eq +H_{12},
\\[0.65ex]
K_{\bar 12}\eq -C_{12},
&\quad&
C_{\bar 12}\eq +K_{12}.
\end{array}
\]
The combinations $\half (A_{12}+B_{12}\pm1)$ and $\half (D_{12}+E_{12}\pm1)$ 
remain invariant. 
This transformation is compensated by 
the map \eqref{eq:stransconj} on the first site of
the fundamental r-matrix in \tabref{tab:rmatrix}.
The transformation for $x_2,\gamma_2$ is the same as above
except that $C,F,H,K$ transform differently
\[
H_{1\bar 2}= +C_{12},\qquad
C_{1\bar 2}= -H_{12},\qquad
K_{1\bar 2}= +F_{12},\qquad
F_{1\bar 2}= -K_{12}.
\]
Under the combined transformation of both sites
the coefficients are invariant up to the following permutations 
\[
C_{\bar 1\bar 2}=F_{12},\qquad 
F_{\bar 1\bar 2}=C_{12},\qquad 
H_{\bar 1\bar 2}=K_{12},\qquad 
K_{\bar 1\bar 2}=H_{12}.
\]

The above transposition map has two fixed points which will be of importance later
\[\label{eq:selfdual}
x^*_\pm=\pm 1\,,
\qquad
z^*_\pm=(ih\pm h')^2\,,
\qquad
q^*=\infty.
\]
These two points will be called \emph{self-dual}.

\subsection{Inversion}
\label{sec:inversion}

Another useful discrete map is the inversion of $z$.
It implies the following transformations of the related parameters
\[
x'= i\,\frac{hx+ih'}{h'x-ih}\,,\qquad
z'= \frac{1}{z}\,.
\]
The parameters of the fundamental representation transform according to
\[
T'= RT,\qquad
q'= zq,\qquad
\gamma'= \frac{\gamma}{h'x-ih}\,,\qquad
R=\matr{cc}{0&i\alpha\\i\alpha^{-1}&0}.
\]
These rules suggest that the following map 
\[
\matr{c}{\gen{Q}^{\prime\,\alpha b}\\\gen{S}^{\prime\,\alpha b}}
= R\matr{c}{\gen{Q}^{\alpha b}\\\gen{S}^{\alpha b}},\qquad
\gencen'= z\gencen,\qquad
\gender'= z^{-1}\gender.
\]
together with $z'=1/z$ is an algebra automorphism. 
Indeed one can confirm that the algebra 
in \secref{sec:classical,sec:affine}
is invariant under the map.
It does not, however, respect the decomposition
in \eqref{eq:decompose} underlying the r-matrix
which is therefore not invariant. 
In particular, the subalgebras $\alg{g}^+$ and $\alg{g}^-$ 
in \eqref{eq:decompose} are interchanged except for the elements 
$\gen{R}^{11},\gen{R}^{22},\gen{L}^{11},\gen{L}^{22}$. 
To achieve a proper transformation we have to interchange them
using the map $z'=1/z$ with
\[\label{eq:inversionalg}
\begin{array}{rcl}
\gen{R}^{\prime\,ab}\eq\varepsilon_{ac}\varepsilon_{bd}\gen{R}^{cd},
\\[0.65ex]
\gen{L}^{\prime\,\alpha\beta}\eq\varepsilon_{\alpha\gamma}\varepsilon_{\beta\delta}\gen{L}^{\gamma\delta},
\end{array}
\qquad
\matr{c}{\gen{Q}^{\prime\,\alpha b}\\\gen{S}^{\prime\,\alpha b}}
= \varepsilon_{\alpha\gamma}\varepsilon_{bd} R\matr{c}{\gen{Q}^{\gamma d}\\\gen{S}^{\gamma d}},\qquad
\begin{array}{rcl}
\gencen'\eq z\gencen,
\\[0.65ex]
\gender'\eq z^{-1}\gender.
\end{array}
\]
Under the inversion all the coefficients $A_{12},\ldots,L_{12}$ for 
the fundamental r-matrix in \tabref{tab:rcoeffs} flip sign.
This yields an overall sign in the r-matrix except for the 
elements $\half(A_{12}+B_{12}\pm 1)$ and $\half(D_{12}+E_{12}\pm 1)$
which are permuted. The permutation is compensated by the 
transformation in \eqref{eq:inversionalg}.
 

\subsection{Statistics Flip}
\label{sec:statistics}

The superalgebras of the kind $\alg{psl}(n|n)$
have an exceptional automorphism \cite{Serganova:1985aa,Grantcharov:2004aa}:
It interchanges the two $\alg{sl}(n)$ factors 
and thus flips the two gradings in certain representations. 
It is responsible for the existence of the two types of
strange superalgebras.

The Lie brackets are invariant under 
the exchange of the two $\alg{sl}(2)$ subalgebras 
\[
\gen{R}^{\prime\,ab}= \gen{L}^{ab},\qquad
\gen{L}^{\prime\,\alpha\beta}= \gen{R}^{\alpha\beta}.
\]
At the level of the fundamental representation
the exchange is compensated by the map
\[\label{eq:BFmap}
\state{\phi^a}'=\state{\psi^a},\qquad
\state{\psi^\alpha}'=\state{\phi^\alpha}.
\]
Under this map the fundamental r-matrix in \tabref{tab:rmatrix} 
flips sign provided that the coefficients 
transform according to 
\[\begin{array}[b]{rclcrclcrclcrclcrcl}
A'_{12}\eq D_{12},&\quad&
B'_{12}\eq E_{12},&\quad&
G'_{12}\eq -L_{12},&\quad&
C'_{12}\eq F_{12},&\quad&
H'_{12}\eq K_{12},
\\[0.65ex]
D'_{12}\eq A_{12},&\quad&
E'_{12}\eq B_{12},&\quad&
L'_{12}\eq -G_{12},&\quad&
F'_{12}\eq C_{12},&\quad&
K'_{12}\eq H_{12}.
\end{array}
\]
Note that the elements $C,F,H,K$ 
in \tabref{tab:rmatrix} receive an extra sign 
due to the change of statistics of the states
when acting with the bifermionic contributions
\eqref{eq:rclass}.
For the coefficients in \tabref{tab:rcoeffs} this transformation is realised 
by mapping the parameter $\gamma$ according to 
\[
\gamma'=\frac{\alpha (h'x-ih)(hx+ih')x}{h'\,\gamma (x^2-1)}\,.
\]
The transformation of the coefficients for the fundamental representation 
in \eqref{eq:abcdclass,eq:matrices} reads
\[
T'=
RT\matr{cc}{0&1\\1&0}
,
\qquad
q'=q,
\qquad
R=\frac{i}{h'}\matr{cc}{+h&+\alpha z^{-1}\\-\alpha^{-1}z&-h}.
\]
The off-diagonal matrix multiplying $T$ corresponds 
to the action \eqref{eq:BFmap}.
The map implies the following transformation for the remaining generators 
\[
\matr{c}{\gen{Q}^{\prime\,\alpha b}\\\gen{S}^{\prime\,\alpha b}}=
R
\matr{c}{\gen{Q}^{\alpha b}\\\gen{S}^{\alpha b}},
\qquad
\gencen'=\gencen,
\qquad
\gender'=-\gender.
\]
This transformation respects the algebra in \secref{sec:classical,sec:affine}
and the decomposition \eqref{eq:decompose} while
it flips the sign of the r-matrix in \eqref{eq:rclassfunct}.

\subsection{Duality}
\label{sec:duality}

Further scrutiny suggests that there is a relationship between 
r-matrices with global parameters $h$ and $h'$ interchanged.
The quadratic relation $h^2+h'^2=1$ implies various sign ambiguities in the map
which we can lift by choosing a different parameter
\[
h=\half(k+k^{-1})\,,\qquad
h'=-\ihalf(k-k^{-1})\,.
\]
The interchange corresponds to the map $k'=ik$.
The coefficients of the fundamental r-matrix in \tabref{tab:rcoeffs} 
turn out to be invariant under the transformation
\[
k'=ik,\qquad
z'=-z,\qquad
x'=x
\,,\qquad
\alpha'=-i\, \frac{k+k^{-1}}{k-k^{-1}}\,\alpha.
\]
The remaining parameters of the fundamental representation transform according to
\[
T'=RT,\qquad
q'=i\,\frac{k-k^{-1}}{k+k^{-1}}\,q,\qquad
\gamma'=\gamma,\qquad
R=\matr{cc}{1&0\\\alpha^{-1}h^{-1}z&1}.
\]
Again the algebra in \secref{sec:classical,sec:affine}
is invariant if one imposes the following map for the generators
\[
\matr{cc}{\gen{Q}^{\prime\,\alpha b}\\\gen{S}^{\prime\,\alpha b}}
=R\matr{cc}{\gen{Q}^{\alpha b}\\\gen{S}^{\alpha b}},
\qquad
\gencen'=i\,\frac{k-k^{-1}}{k+k^{-1}}\,\gencen,
\qquad
\gender'=-i\,\frac{k+k^{-1}}{k-k^{-1}}\,\gender.
\]
Also the decomposition \eqref{eq:decompose} is respected, 
and consequently the r-matrix is invariant.
In particular, the coefficients in \tabref{tab:rcoeffs} transform trivially.

%

\subsection{Reparametrisation}
\label{sec:uniform}

Here we introduce a change of variables
which helps to make some features of the algebra
discussed above somewhat more transparent. 
This will be instructive to some extent, 
but in the remainder of the paper we shall nevertheless stick 
to the old variables.

\paragraph{Reparametrisation.}

We have seen in \secref{sec:conjugation} 
that for each value of $z$ there are four
fundamental representations. 
They are distinguished by different values of $x$ and $\gamma$.
For instance, for each $z$ the map 
\eqref{eq:zq} permits two values for $x$,
and for each $x$ there is a pair of 
representations distinguished by different signs for $\gamma$. 

In fact one can introduce a new parameter $y$ to distinguish 
all four fundamental representations corresponding to a
particular value $z$
\[
 x=-\frac{y^2-1}{y^2+1}\,,
\quad
\gamma=\frac{y^2+k^2}{2ky}\,\eta,
\quad
z=-k^2\,\frac{y^4-1}{y^4-k^4}\,,
\quad
q=\frac{-1}{k^2(k-k^{-1})}\,\frac{y^4-k^4}{y^2}\,.
\]
At the same time we shall use the parameter $k$ introduced in 
\secref{sec:duality} instead of $h$
\[
h=\half(k+k^{-1}),
\qquad
h'=-\ihalf(k-k^{-1}),
\qquad
\alpha=\half(k-k^{-1})\kappa.
\]
Altogether the following parametrisation yields a slightly more transparent picture.
This can be observed for the coefficients $a,b,c,d$ of the matrix $T$ 
which now take a very symmetric form
\[
a=\frac{\eta(y^2+k^2)}{2ky}\,,
\quad
b=\frac{-\kappa(y^2-k^2)}{2k\eta y}\,,
\quad
c=\frac{-\eta(y^2+1)}{\kappa (k-k^{-1})y}\,,
\quad
d=\frac{y^2-1}{\eta(k-k^{-1})y}\,.
\]
Notably, all the coefficients in \tabref{tab:rcoeffs} 
now factor completely into terms $y^2\pm 1$, $y^2\pm k^2$
and $y_1^2\pm y_2^2$.

\paragraph{Special Points.}

\begin{figure}
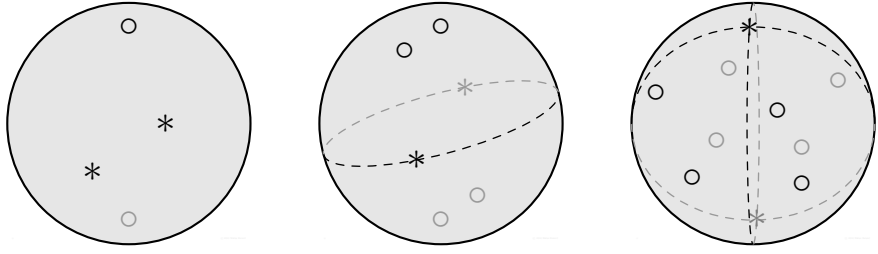
\centering
\includegraphics{FigSphereZ.mps}\qquad
\includegraphics{FigSphereX.mps}\qquad
\includegraphics{FigSphereY.mps}%
\caption{The compactified complex plane for $z$, $x$ or $y$, respectively. 
The points corresponding to $z^\circ_\pm=0,\infty$ are marked by $\circ$. 
The self-dual points corresponding to $z^\ast_\pm$ are marked by $\ast$.
The spheres are divided into one, two or four regions 
which are identified by a twist.}
\label{fig:special}
\end{figure}

Investigating the above expressions it becomes
clear that the $y$-plane has several special points:
The points $y^\circ_{+1,2,3,4}=\pm 1,\pm i$ map to $z=0$
while $y^\circ_{-1,2,3,4}=\pm k,\pm ik$ map to $z=\infty$. 
Finally, the two points $y^*_{\pm}=0,\infty$ map to the self-dual 
points $z^*_\pm=-k^{\mp 2}$ or $x^*_\pm=\pm 1$ in \eqref{eq:selfdual}.
The configurations of special points are displayed in \figref{fig:special}.

\paragraph{Eigenbasis.}

A curious feature of the matrix $U$ in \eqref{eq:UVmatrix}
is that the $z$-dependence is in the prefactor only. 
Hence the eigenvectors are constants and we 
can use them as a new basis for $\gen{Q}$ and $\gen{S}$. 

A matrix to perform the similarity transformation to the 
eigenvectors is given by 
\[
R=\matr{cc}{-2/(k-k^{-1})&\kappa k\\-\kappa^{-1}/(k-k^{-1})&\half k^{-1}}.
\]
The resulting matrix $\tilde T$ containing the coefficients 
$\tilde a,\tilde b,\tilde c,\tilde d$ then reads simply
\[
\tilde T=RT=\matr{cc}{\eta y&\kappa \eta^{-1}y\\-\half \kappa^{-1}\eta y^{-1}&\half\eta^{-1}y^{-1}}.
\]
The transformation curiously removes the diagonal terms in the matrix $\tilde W$ 
\[
\tilde W=RWR^{-1}=\frac{1}{k+k^{-1}}
\matr{cc}{0&-4\kappa (1+k^2z)\\\kappa^{-1}(1+k^{-2}z)&0},
\]
whereas by construction the matrix $\tilde U$ is diagonal
\[
\tilde U=RUR^{-1}=\frac{(k^4-1)z}{4(k^2+z)(1+k^2z)}
\matr{cc}{+1&0\\0&-1}.
\]
The only reason not to perform this similarity transformation once and for all
is that it obscures the linear combinations 
of $\tilde{\gen{Q}}$ and $\tilde{\gen{S}}$ which 
appear in the contribution \eqref{eq:tr0} to the r-matrix 
and in the triangular decomposition \eqref{eq:decompose}.
We will thus stick to the original basis of $\gen{Q}$ and $\gen{S}$.

\paragraph{Embedding.}

The above reparametrisation has led to rational
expressions for the parameters $a,b,c,d,q$ of the fundamental representation.%
\footnote{In fact, also $x$ (but not $z$ itself) 
is permissible because $T(x),q(x)$ are rational.}
We can use them to go one step further, and embed our algebra into the 
standard algebra $\alg{gl}(2|2)[y,y^{-1}]$ 
(with $\bar W=M$, $\bar U=\bar V=0$)
in analogy to the transformation in \cite{Beisert:2007ty}
\[
\begin{array}{rcl}
\gen{R}^{ab}\eq \bar{\gen{R}}^{ab},
\\[0.65ex]
\gen{L}^{\alpha\beta}\eq \bar{\gen{L}}^{\alpha\beta},
\end{array}
\quad
\matr{c}{\gen{Q}^{\alpha b}\\\gen{S}^{\alpha b}}=
T(y)\matr{c}{\bar{\gen{Q}}^{\alpha b}\\\bar{\gen{S}}^{\alpha b}},
\quad
\begin{array}{rcl}
\gencen\eq q(y)\bar{\gencen},
\\[0.65ex]
\gender\eq q(y)^{-1}\bar{\gender},
\end{array}
\]
with $\eta=\sqrt{\kappa}$.
Note that one must allow for pole singularities 
at the special points $y^\circ,y^*$.
In this sense, one has to require
that the Riemann surface underlying 
the ambient algebra is a sphere with punctures 
at all of these points, see \figref{fig:special}, 
not just at $y=0,\infty$ as for conventional loop algebras.

The reduction to our subalgebra is done 
by twisting with the $\Integers_4$-periodic automorphism of $\alg{gl}(2|2)$
\footnote{Although the automorphism is $\Integers_4$-periodic, 
it merely corresponds to a $\Integers_2$-periodic \emph{outer} automorphism 
of the $\alg{gl}(2|2)$ algebra,
see the discussion below \protect\eqref{eq:aboveZ2}.}
\[
y\to iy,
\quad
\bar{\gen{Q}}\to i\bar{\gen{S}},
\quad
\bar{\gen{S}}\to i\bar{\gen{Q}},
\quad
\bar{\gencen}\to -\bar{\gencen},
\quad
\bar{\gender}\to -\bar{\gender}.
\]
Furthermore, singularities at the fixed points $y^*=0,\infty$
are restricted to be at most double poles while 
there can be poles of arbitrary order at the points $y^\circ$. 

As above, this redefinition changes the form of the r-matrix \eqref{eq:rclassfunct} 
and the triangular decomposition \eqref{eq:decompose},
and we shall refrain from making use of it subsequently.
It is nevertheless interesting because it
shifts the deformation from the algebra 
to the r-matrix,
i.e.\ the conventional affine $\alg{gl}(2|2)$ algebra apparently
admits a non-standard r-matrix.

\paragraph{Discrete Transformations.}

The discrete transformations discussed above also simplify:
Essentially they map the various special points $y^\circ$ and $y^*$
into each other.
The conjugation symmetry discussed in \secref{sec:conjugation}
translates between the four conjugate fundamental representations
for each value of $z$. This is achieved through
\[
y'=iy,\qquad
\eta'=\frac{i\kappa}{\eta}\,.
\]
The inversion symmetry discussed in \secref{sec:inversion}
is invoked by
\[
y\to \frac{k}{y}\,,\qquad
\eta'=i\eta.
\]
The statistics flip symmetry in \secref{sec:statistics} requires
to change $\eta$ according to 
\[
\eta'=\frac{\kappa}{\eta}\,\frac{y^2-1}{y^2+1}\,\frac{y^2-k^2}{y^2+k^2}\,.
\]
Finally, there is the duality discussed in \secref{sec:duality}
which relates 4 different values of $k$ 
\[
k'=ik,\qquad
\eta'=i\,\frac{y^2+k^2}{y^2-k^2}\,\eta,\qquad
\kappa'=-\kappa.
\]
A similar transformation does not change anything in the original
parametrisation
\[
k'=\frac{1}{k}\,,\qquad
y'=\frac{1}{y}\,,\qquad
\kappa'=-\kappa.
\]
Note that the point $k=\sqrt{i}$ is self-dual under a combination 
of the above two duality maps. 
This map thus becomes an additional symmetry of the $k=\sqrt{i}$ system
\[
y'=\frac{1}{y}\,,\qquad
\eta'=-i\,\frac{y^2+i}{y^2-i}\,\eta,\qquad
x'=-x,\qquad
z'=-z.
\]
It might be worth investigating if the self-dual point $k=\sqrt{i}$ 
has further interesting properties.
The other self-dual point $k=1$ is 
discussed in the following section.

\section{Limits}
\label{sec:limits}

The r-matrix presented in \tabref{tab:rmatrix,tab:rcoeffs}
has a couple of interesting limits which themselves
lead to quasi-triangular Lie algebras.
We shall call the r-matrix of \secref{sec:classical}
the ``full trigonometric r-matrix''.
The limits will modify the attributes of the name accordingly.

\subsection{Full Rational Case}
\label{sec:rational}

The trigonometric r-matrix can be reduced to the 
rational r-matrix \cite{Klose:2006zd,Torrielli:2007mc} 
obtained in the context of the AdS/CFT duality. 
To that end one takes the limit
\[\label{eq:ratlimit}
h=\epsilon\to 0,\qquad
x\sim \epsilon^0,\qquad
z=1+i\epsilon u+\order{\epsilon^2}.
\]
All of the following results are in full agreement with \cite{Beisert:2007ty}
where the structure and the underlying quasi-triangular Lie bialgebra 
were obtained.

\paragraph{Fundamental r-Matrix.}

The parameters of the fundamental representation 
\eqref{eq:abcdclass,eq:zq} become%
\footnote{For convenience, 
one might absorb several factors of $i$ into the
definition of $q,\gencen,\gender,\alpha$.}
\[
a=\gamma,\quad
b=\frac{-i\alpha x}{\gamma(x^2-1)}\,,\quad
c=\frac{i\gamma}{\alpha x}\,,\quad
d=\frac{x^2}{\gamma(x^2-1)}\,,\quad
u=x+\frac{1}{x}\,,\quad
q=\frac{-i\coni x}{x^2-1}\,.
\]
In this limit the r-matrix diverges like $\epsilon^{-1}$ and needs 
to be renormalised
\[
\tilde r=i\epsilon r.
\]
Most importantly, the divergence reduces 
the structure of the r-matrix in \tabref{tab:rmatrix}
because the constant terms in the 
combinations $\half (A+B\pm 1)$ and $\half (D+E\pm 1)$ drop out. 
It can then be written in a manifestly $\alg{sl}(2)\oplus\alg{sl}(2)$ 
invariant fashion known for \emph{rational} r-matrices
\<\label{eq:rrational}
\tilde r\state{\phi^a\phi^b}\eq
\half (\tilde A_{12}+\tilde B_{12})\state{\phi^b\phi^a}
+\half (\tilde A_{12}-\tilde B_{12})\state{\phi^a\phi^b}
+\half \tilde C_{12} \varepsilon^{ab}\varepsilon_{\gamma\delta}
\state{\psi^\gamma\psi^\delta},
\nln
\tilde r\state{\psi^\alpha\psi^\beta}\eq
-\half (\tilde D_{12}+\tilde E_{12})\state{\psi^\beta\psi^\alpha}
-\half (\tilde D_{12}-\tilde E_{12})\state{\psi^\alpha\psi^\beta}
-\half \tilde F_{12}\varepsilon^{\alpha\beta}\varepsilon_{cd}\state{\phi^c\phi^d},
\nln
\tilde r\state{\phi^a\psi^\beta}\eq
\tilde G_{12}\state{\phi^a\psi^\beta}
+\tilde H_{12}\state{\psi^\beta\phi^a},
\nln
\tilde r\state{\psi^\alpha\phi^b}\eq
\tilde K_{12}\state{\phi^b\psi^\alpha}
+\tilde L_{12}\state{\psi^\alpha\phi^b}.
\>
The coefficient functions $\tilde A,\ldots,\tilde L$ are essentially
the same as $A,\ldots,L$ in \tabref{tab:rcoeffs},
but the $z$-dependence reduces according to the limit
\[
\frac{z_1}{z_1-z_2}
\sim
\frac{\half z_1+\half z_2}{z_1-z_2}
\sim
\frac{z_2}{z_1-z_2}
\to
\frac{1}{i\epsilon}\,\frac{1}{u_1-u_2}\,,
\]
where the factor of $1/i\epsilon$ is absorbed into the definition of the 
r-matrix $\tilde r$.
The coefficients obey the same linear relations \eqref{eq:rellin}
as in the trigonometric case, 
but the constant shift disappears from the 
quadratic relations \eqref{eq:relquad}
\[
\quarter(\tilde A+\tilde B)(\tilde D+\tilde E)
=
\sfrac{1}{4}(3\tilde A-\tilde B)(3\tilde D-\tilde E)+4\tilde G\tilde L
=\tilde C\tilde F+\tilde H\tilde K.
\]

\paragraph{Algebra and Universal r-Matrix.}

The loop algebra derived in \secref{sec:loops}
remains essentially the same. 
One difference is that we shall use $u$ as the formal loop variable
instead of $z$. Thus the brackets in \eqref{eq:DQmatrix}
now read
\<\label{eq:DQmatrixrat}
\acomm{\gen{Q}^{\alpha b}}{\gen{Q}^{\gamma d}}
\eq \varepsilon^{\alpha\gamma}\varepsilon^{bd} W_{12}(u)\,\gencen,
\nln
\acomm{\gen{Q}^{\alpha b}}{\gen{S}^{\gamma d}}
\eq 
-\varepsilon^{\alpha\gamma}\gen{R}^{bd}
+\varepsilon^{bd}\gen{L}^{\alpha\gamma}
-\varepsilon^{\alpha\gamma}\varepsilon^{bd}W_{11}(u)\, \gencen,
\nln
\acomm{\gen{S}^{\alpha b}}{\gen{S}^{\gamma d}}
\eq -\varepsilon^{\alpha\gamma}\varepsilon^{bd} W_{21}(u)\,\gencen,
\nln
\comm{\gender}{\gen{Q}^{\alpha b}}
\eq
W_{11}(u)\,\gen{Q}^{\alpha b}
+W_{12}(u)\,\gen{S}^{\alpha b},
\nln
\comm{\gender}{\gen{S}^{\alpha b}}
\eq
W_{21}(u)\,\gen{Q}^{\alpha b}
+W_{22}(u)\,\gen{S}^{\alpha b}.
\>
Furthermore the matrix $W$ in \eqref{eq:matrices}
reduces to%
\footnote{The generators $\gender,\gencen$ are shifted
by one loop level w.r.t.\ the corresponding ones in 
\protect\cite{Beisert:2007ty},
i.e.\ $W$ differs by a factor of $u$.}
\[
W=\conw \matr{cc}{+i u&2\alpha\\2\alpha^{-1}&-i u}.
\]

The limit of the universal trigonometric r-matrix in \eqref{eq:rclassfunct}
reads
\[
\tilde r=\frac{1}{u_1-u_2}\,\tilde t.
\]
When expanded into loop levels using a geometric series (cf.\ \secref{sec:levels}) 
one finds the analog of \eqref{eq:rclass}
\<
\tilde r\eq 
+\sum_{k=0}^\infty \Big[
-\varepsilon_{ac}\varepsilon_{bd}\gen{R}^{ab}_{-1-k}\otimes \gen{R}^{cd}_{+k}
+\varepsilon_{\alpha\gamma}\varepsilon_{\beta\delta}\gen{L}^{\alpha\beta}_{-1-k}\otimes \gen{L}^{\gamma\delta}_{+k}
\nl\qquad\qquad
-\varepsilon_{\alpha\gamma}\varepsilon_{bd}\gen{Q}^{\alpha b}_{-1-k}\otimes \gen{S}^{\gamma d}_{+k}
+\varepsilon_{\alpha\gamma}\varepsilon_{bd}\gen{S}^{\alpha b}_{-1-k}\otimes \gen{Q}^{\gamma d}_{+k}
\nl\qquad\qquad
-\gencen_{-1-k}\otimes \gender_{+k}
-\gender_{-1-k}\otimes \gencen_{+k}
\Big].
\>

\paragraph{Affine Extension.}

The loop variable $z$ is replaced by $u$ according to \eqref{eq:ratlimit}
\[
z=1+i\epsilon u+\order{h^2}.
\]
After a rescaling
\[
\tilde{\genaffder}=i\epsilon\genaffder
\]
the affine derivation \eqref{eq:affderloops}
transforms into a derivative w.r.t.\ $u$
\[
\comm{\tilde{\genaffder}}{u}=1
\quad
\mbox{or}
\quad
\tilde{\genaffder}\simeq \frac{d}{du}\,.
\]
The structure of the affine derivations remains the same
as in \eqref{eq:DQaffder}
\<\label{eq:DQaffderrat}
\comm{\tilde{\genaffder}}{\gen{Q}^{\alpha b}}
\eq
\tilde{U}_{11}(u)\,\gen{Q}^{\alpha b}
+\tilde{U}_{12}(u)\,\gen{S}^{\alpha b},
\nln
\comm{\tilde{\genaffder}}{\gen{S}^{\alpha b}}
\eq
\tilde{U}_{21}(u)\,\gen{Q}^{\alpha b}
+\tilde{U}_{22}(u)\,\gen{S}^{\alpha b},
\nln
\comm{\tilde{\genaffder}}{\gencen}
\eq +\tilde{V}(u)\,\gencen,
\nln
\comm{\tilde{\genaffder}}{\gender}
\eq -\tilde{V}(u)\,\gender,
\>
whereas for the central charges in \eqref{eq:DQcentral} 
one has to replace $dz/z$ by $du$
%
\<\label{eq:DQcentralrat}
\bigacomm{f(u)\gen{Q}^{\alpha b}}{g(u)\gen{Q}^{\gamma d}}
\eq 
\varepsilon^{\alpha\gamma}\varepsilon^{bd} f(u)g(u)W_{12}(u)\, \gencen
\nl
+\varepsilon^{\alpha\gamma}\varepsilon^{bd}
\frac{1}{2\pi i}\oint \bigbrk{f(u)g(u)U_{12}(u)\,du}\,\genaffcen,
\nln
\bigacomm{f(u)\gen{Q}^{\alpha b}}{g(u)\gen{S}^{\gamma d}}
\eq
f(u)g(u)
\lrbrk{ 
-\varepsilon^{\alpha\gamma}\gen{R}^{bd}
+\varepsilon^{bd}\gen{L}^{\alpha\gamma}
-\varepsilon^{\alpha\gamma}\varepsilon^{bd}W_{11}(u)\, \gencen
}
\nl
+\varepsilon^{\alpha\gamma}\varepsilon^{bd}
\frac{1}{2\pi i}\oint 
\bigbrk{f(u)dg(u)-f(u)g(u)U_{11}(u)\,du}
\,\genaffcen,
\nln
\bigacomm{f(u)\gen{S}^{\alpha b}}{g(u)\gen{S}^{\gamma d}}
\eq 
-\varepsilon^{\alpha\gamma}\varepsilon^{bd}f(u)g(u) W_{21}(u)\, \gencen
\nl
+\varepsilon^{\alpha\gamma}\varepsilon^{bd}
\frac{1}{2\pi i}\oint 
\bigbrk{-f(u)g(u)U_{21}(u)\,du}
\,\genaffcen,
\nln
\bigcomm{f(u)\gencen}{g(u)\gender}
\eq 
-\frac{1}{2\pi i}\oint 
\bigbrk{f(u)dg(u)-V(u) f(u)g(u)\,du}
\,\genaffcen.
\>
The parameters have to be rescaled w.r.t.\ \eqref{eq:UVmatrix},
now they read, see also \cite{Young:2007wd},
\[
\tilde U=i\epsilon U=
\frac{1}{u^2-4}\matr{cc}
{0&-i\alpha\\+i\alpha^{-1}&0},
\qquad
\tilde V=i\epsilon V=
-\frac{u}{u^2-4}\,.
\]

Note that the non-vanishing of the above parameters
leads to the non-invariance of the r-matrix under $\tilde{\genaffder}$
and thus to a non-trivial cobracket (see \secref{sec:affine}).
When $\tilde{\genaffder}$ is interpreted as a two-dimensional (Lorentz) boost,
the corresponding (Lorentz) symmetry would be deformed 
along the lines discussed in \cite{Gomez:2007zr,Young:2007wd}.

It appears that the exponentiated affine derivation
$\exp(\ihalf g^{-1}\tilde\genaffder)$
(note that exponentiated generators naturally appear in quantum algebras,
see \secref{sec:quantum})
plays an important role in the quantisation of the algebra:
The generator induces a finite shift of $u$ by an amount
which frequently occurs in the quantum R-matrix, 
e.g.\ 
\[
x^\pm(u)=x(u\pm \ihalf g^{-1})=
\exp(\pm\ihalf g^{-1}\tilde\genaffder)\,x(u). 
\]
It would be interesting to pursue the role of the affine derivation further.

\subsection{Conventional Rational Case}
\label{sec:ratconv}

The simplest limit of the fundamental r-matrix 
is obtained when the two 
parameters $x_k$ approach each other at a generic point $x_0$
\[
x=x_0(1+\epsilon u).
\]
The r-matrix diverges in the limit $\epsilon\to 0$
and one obtains a rational r-matrix $\tilde r$ \eqref{eq:rrational}
\[\label{eq:genericlimit}
\tilde r
=\frac{-hh'\epsilon(x_0^2-1)}{(hx_0+ih')(h'x_0-ih)}\,r.
\]
The new coefficient functions all have the same simple singularity
at $u_1=u_2$
\[
\tilde A_{12}=\tilde B_{12}=\tilde D_{12}=\tilde E_{12}=
\frac{\gamma_2}{\gamma_1}\tilde H_{12}=
\frac{\gamma_1}{\gamma_2}\tilde K_{12}=
\frac{1}{u_1-u_2}\,,
\quad
\tilde C_{12}=\tilde F_{12}=
\tilde G_{12}=\tilde L_{12}=
0.
\]
This r-matrix is the fundamental representation of the
classical rational r-matrix for 
the conventional affine $\alg{gl}(2|2)$ algebra.

Interestingly, the parameter $h$ has dropped out
completely from the r-matrix $\tilde r$
and from the associated affine bialgebra. 
However this does not mean that the limit is the same for all
$h$ and for all $x_0$. In particular one can see that the prefactor
in \eqref{eq:genericlimit} is singular at certain points $x_0$,
namely $x_0=-ih'/h$, $x_0=ih/h'$ and $x_0=\pm 1$.
The first pair of points corresponds to $z_0=\infty$
and the second pair to the self-dual points $z_0=z^\ast_\pm=(ih\pm h')^2$
discussed around \eqref{eq:selfdual}.
In the following we shall discuss the limits at these points.

\subsection{Conventional Trigonometric Case}
\label{sec:trigconv}

Let us next discuss the point $z_0=\infty$.
The point $z_0=0$ is analogous according to the 
discussion in \secref{sec:inversion}, and there is no 
need to discuss it separately.
Similarly, we can safely restrict to one of the two corresponding points
$x=ih/h'$ and $x=-ih'/h$, cf.\ \secref{sec:conjugation}.
Here we take the limit
\[
x=\frac{ih}{h'}\lrbrk{1+\frac{\epsilon}{\tilde z}+\order{\epsilon^2}},
\qquad
z=\epsilon^{-1}\tilde z.
\]
At the same time, the parameter $\alpha$ 
should scale like $\alpha\sim \epsilon^{-1}$.
In this case the r-matrix remains finite 
in the limit $\epsilon\to 0$.
Thus the trigonometric structure in \tabref{tab:rmatrix} applies, 
and its coefficients in \tabref{tab:rcoeffs} reduce to
\[\label{eq:stdtrigres}
\begin{array}[b]{c}
\displaystyle 
A_{12}=B_{12}=D_{12}=E_{12}=
\frac{\half \tilde z_1+\half \tilde z_2}{\tilde z_1-\tilde z_2}\,,
\quad
C_{12}=F_{12}=0,
\\[2ex]
\displaystyle 
G_{12}=-L_{12}=
+\frac{1}{4}\,,
\quad
H_{12}=
\frac{\gamma_1}{\gamma_2}\,
\frac{\tilde z_1}{\tilde z_1-\tilde z_2}\,,
\quad
K_{12}=
\frac{\gamma_2}{\gamma_1}\,
\frac{\tilde z_2}{\tilde z_1-\tilde z_2}\,.
\end{array}
\]
These coefficients are precisely the coefficients of the conventional
trigonometric r-matrix for $\alg{gl}(2|2)$.
The underlying algebraic structure is thus the standard affine $\alg{gl}(2|2)$
algebra with trigonometric r-matrix.



\subsection{Twisted Rational Case}
\label{sec:rattwist}

The self-dual points $z^*_\pm$ lead to 
a more elaborate limit. 
According to \secref{sec:inversion} the two limits 
are equivalent and we choose to investigate 
\[
z_0=(ih+h')^2,\qquad
x_0=+1.
\]
The limit is defined by 
\[
x=1+\epsilon\,\frac{h'-ih}{h'}\,y,
\qquad
z=z_0\lrbrk{1+\frac{ih\epsilon^2 }{h'}\,u},
\qquad
u=y^2,
\qquad
\alpha=\frac{i\epsilon}{h'-ih}\,\tilde\alpha.
\]
Here the r-matrix diverges quadratically
\[
\tilde r=\frac{ih\epsilon^2}{h'}\,r
\]
with $\tilde r$ a rational r-matrix of the form \eqref{eq:rrational}.
The coefficients of this fundamental r-matrix read
\<\label{eq:rtwistrat}
\tilde A_{12}=\tilde D_{12}\eq \frac{1}{4y_1y_2}\,\frac{y_1+y_2}{y_1-y_2}\,,
\nln
\half(\tilde A_{12}+\tilde B_{12})=
\half(\tilde D_{12}+\tilde E_{12})
\eq \frac{1}{y_1^2-y_2^2}\,,
\nln
\half(\tilde A_{12}-\tilde B_{12})=
\half(\tilde D_{12}-\tilde E_{12})
\eq 
\frac{1}{4y_1y_2}\,\frac{y_1-y_2}{y_1+y_2}\,,
\nln
\frac{\tilde\alpha}{2\gamma_1\gamma_2}\,\tilde C_{12}=
-\frac{2\gamma_1\gamma_2y_1y_2}{\tilde\alpha}\,\tilde F_{12}\eq 
\frac{1}{2}\,\frac{1}{y_1+y_2}\,,
\nln
\tilde G_{12}=-\tilde L_{12}\eq \frac{1}{4y_1y_2}\,,
\nln
\frac{\gamma_2 y_2}{\gamma_1}\,\tilde H_{12}=
\frac{\gamma_1 y_1}{\gamma_2}\,\tilde K_{12}\eq 
\frac{1}{2}\,\frac{1}{y_1-y_2}\,.
\>
These coefficients along with the rational r-matrix structure in \eqref{eq:rrational}
agree with Eqs.\ (4.2, 4.10) in \cite{Klose:2007rz}
when setting $y=1/2p_-$, $\gamma=\sqrt{\tilde\alpha p_-}$.
This case therefore provides the classical s-matrix 
for strings in $AdS_5\times S^5$ 
in the near flat space limit \cite{Maldacena:2006rv}.

In order to understand the algebra underlying this r-matrix, 
we consider the coefficients \eqref{eq:abcdclass,eq:zq} for the fundamental
representation first. It turns out that $a$ and $b$ are finite
while $c$ and $d$ diverge. In combination with $a$ and $b$ 
one can nevertheless find finite combinations $\tilde c$ and $\tilde d$
\[\label{eq:abcdtwistrat}
a= \gamma,
\quad
b= \frac{\tilde \alpha }{2y\gamma}\,,
\quad
\tilde c=c-\frac{a}{\tilde\alpha\epsilon}= 
-\frac{\gamma y}{\tilde \alpha}
\,,
\quad
\tilde d=d-\frac{b}{\tilde\alpha\epsilon}= \frac{1}{2\gamma}
\,,
\quad
\tilde q=\frac{i\epsilon}{x_0}\,q=
\coni\frac{1}{2y}\,.
\]
In the matrix notation \eqref{eq:matrices} this corresponds to 
a multiplication by a matrix $R$
\[
\tilde T=RT,\qquad
R=\matr{cc}{1&0\\\tilde\alpha^{-1}\epsilon^{-1}&1}.
\]
This implies that we should consider the following redefined generators 
\[
\matr{cc}{\tilde{\gen{Q}}^{\alpha b}\\\tilde{\gen{S}}^{\alpha b}}
=R\matr{cc}{\gen{Q}^{\alpha b}\\\gen{S}^{\alpha b}},
\qquad
\tilde{\gencen}=\frac{i\epsilon}{x_0}\,\gencen,
\qquad
\tilde{\gender}=-\frac{ix_0}{\epsilon}\,\gender.
\]
They have a well-defined algebra in the limit $\epsilon\to 0$, 
cf.\ \eqref{eq:DQmatrix,eq:matrices}
with the new matrix
\[
\tilde W=\frac{x_0}{i\epsilon}RWR^{-1}=\conw \matr{cc}{0&2\tilde\alpha\\2\tilde\alpha^{-1}u&0}.
\]

Next we consider the limit of the affine extension of the algebra. 
The affine derivation must be rescaled
\[
\tilde{\genaffder}=\frac{ih\epsilon^2}{h'}\,\genaffder\simeq \frac{d}{du}\,.
\]
The action on the generators is defined by \eqref{eq:DQaffderrat} 
with coefficients $\tilde U,\tilde V$ \eqref{eq:UVmatrix} limiting to
\[
\tilde U=
\frac{ih\epsilon^2}{h'}\,RUR^{-1}=
\frac{1}{4u}\matr{cc}{-1&0\\0&+1},
\qquad
\tilde V=
\frac{ih\epsilon^2}{h'}\,V=
-\frac{1}{2u}\,.
\]

The above algebra is in fact a twisted affine algebra:
This can be observed if we write the Lie brackets
in terms of the generators
\[\label{eq:twistgen}
\bar{\gen{Q}}=u^{+1/4}\tilde{\gen{Q}},\qquad
\bar{\gen{S}}=u^{-1/4}\tilde{\gen{S}},\qquad
\bar{\gencen}=u^{+1/2}\tilde{\gencen},\qquad
\bar{\gender}=u^{-1/2}\tilde{\gender}.
\]
Now the above algebra is defined by the parameters 
\[\label{eq:aboveZ2}
\bar W=\conw \matr{cc}{0&2\tilde\alpha\\2\tilde\alpha^{-1}&0},
\qquad
\bar U=\bar V=0.
\]
I.e.\ the loop levels of the generators add up simply,
and the affine extension acts canonically.
The automorphism defining the above twist has a period of 4. 
It corresponds to an \emph{outer} $\Integers_2$-automorphism of $\alg{gl}(2|2)$
which acts non-trivially on one of the two $\alg{sl}(2)$ subalgebra.
Note that for the simple superalgebra $\alg{psl}(2|2)$ 
the corresponding automorphism is inner \cite{Serganova:1985aa,Grantcharov:2004aa}, 
so the non-triviality of the twist is 
only due to the central charge $\gencen$ and the derivation $\gender$.

\subsection{Twisted Trigonometric Case}
\label{sec:trigtwist}

We have exhausted all the special values of $z$
for generic values of the global parameter $h$. 
As $h$ varies also the self-dual points \eqref{eq:selfdual}
\[
z^*_\pm=\bigbrk{ih\pm h'}^2
\]
move around in the complex plane
while the special values $z=0$ and $z=\infty$ remain fixed.
For particular values of $h$, namely $h=0,\pm1,\infty$, 
some of the special values coincide 
giving rise to further limits of interest.
We have seen in \secref{sec:duality} that the points $h=\pm 1$ 
are equivalent to $h=0$, consequently there is no need to discuss them 
separately.

Let us first consider the case $h=\infty$. 
There both self-dual points approach the other two special values, 
$z^*_+=0$, $z^*_-=\infty$.
Now we take the limit
\[
h=\epsilon^{-1},\qquad
z\sim \epsilon^0,\qquad
\epsilon\to 0.
\]
Here the parameters \eqref{eq:abcdclass,eq:zq} of the fundamental representation 
\eqref{eq:loopeval} read
\[\label{eq:abcdtwisttrig}
a=\gamma,\quad
b=\frac{\alpha}{2\gamma y}\,,\quad
c=-\frac{\gamma y}{\alpha}\,,\quad
d=\frac{1}{2\gamma}\,,\quad
x=1-\frac{\epsilon}{y}\,,\quad
z=y^2,\quad
q=\coni\frac{1}{2y}\,,
\]
and the fundamental classical r-matrix in \tabref{tab:rmatrix,tab:rcoeffs}
takes the same form using these simplified parameters $a,b,c,d,q$.
In particular we find
\<\label{eq:rtwisttrig}
A_{12}
=D_{12}
\eq
\frac{1}{4}\, \frac{y_1+y_2}{y_1-y_2}\,,
\nln
\half(A_{12}+B_{12}+1)
=\half(D_{12}+E_{12}+1)
\eq \frac{y_1^2}{y_1^2-y_2^2}\,,
\nln
\half(A_{12}+B_{12}-1)
=\half(D_{12}+E_{12}-1)
\eq \frac{y_2^2}{y_1^2-y_2^2}\,,
\nln
\half(A_{12}-B_{12})
=\half(D_{12}-E_{12})
\eq 
-\frac{1}{4}\,\frac{y_1-y_2}{y_1+y_2}\,,
\nln
\frac{\alpha}{2\gamma_1 \gamma_2 y_1y_2}\,
C_{12}=
-\frac{2\gamma_1\gamma_2 }{\alpha}\,
F_{12}\eq
-\frac{1}{2}\,\frac{1}{y_1+y_2}\,,
\nln
G_{12}
=L_{12}
\eq
0,
\nln
\frac{\gamma_2}{y_1\gamma_1}\, 
H_{12}=
\frac{\gamma_1}{y_2\gamma_2}\, 
K_{12}\eq
\frac{1}{2}\,\frac{1}{y_1-y_2}\,.
\>
These coefficients are reminiscent of those 
for the twisted rational r-matrix in \eqref{eq:rtwistrat}.
In fact, the representation parameters in \eqref{eq:abcdtwisttrig} 
agree precisely with those in \eqref{eq:abcdtwistrat}.
Effectively, it means that the two algebras are equivalent
(up to the affine extensions).
The parameters therefore read
\[
W=\conw \matr{cc}{0&2\alpha\\2\alpha^{-1}z&0},
\qquad
U=\frac{1}{4}\matr{cc}{-1&0\\0&+1},
\qquad
V=-\frac{1}{2}\,.
\]
As explained in \secref{sec:rattwist}, 
they describe a $\Integers_2$-twisted affine $\alg{gl}(2|2)$ algebra. 
A quantum R-matrix for this algebra was derived in \cite{Gould:1996aa}.
Therefore one would expect that its classical limit 
is related to the trigonometric r-matrix described above.

It is curious to observe that the coefficients
in \eqref{eq:rtwisttrig} match almost exactly
with those found for the scattering matrix derived in 
(4.9--4.11) in \cite{Hoare:2009fs} 
including the functional form of its prefactor 
when equating $y_k=\exp(\theta_k)$.
Nevertheless, there is a crucial difference: 
The s-matrix in \cite{Hoare:2009fs} is based on the rational structure \eqref{eq:rrational} 
with manifest $\alg{su}(2)\oplus\alg{su}(2)$ symmetry, 
while the coefficients are intimately associated 
to the the trigonometric structure in \tabref{tab:rmatrix} with 
broken $\alg{su}(2)\oplus\alg{su}(2)$.
Effectively $K_2$ in \cite{Hoare:2009fs}
compares to $\half (A_{12}+B_{12})$ rather than 
$\half (A_{12}+B_{12}\pm 1)$.

\subsection{Special Trigonometric Case at \texorpdfstring{$h=\infty$}{h=inf}}
\label{sec:triginf}

The value $h=\infty$ considered above is subtle,
and the result depends on the details
of taking the limit $h\to\infty$. 
Previously we have assumed that $z$ remains finite,
but there is also the option of scaling $z\to 0$ or $z\to\infty$ 
in correlation with $h\to\infty$. 
Moreover the result generally depends on how fast $z$ converges in
comparison to $h$. 
A suitable limit with $z\to 0$ 
(according to \secref{sec:inversion} this is equivalent to $z\to\infty$)
turns out to be 
\[
h=\epsilon^{-1},\qquad
z=-\quarter\epsilon^2\tilde z,\qquad
x\sim\epsilon^0,\qquad
\alpha=\half\epsilon\tilde\alpha.
\]
This limit is distinguished from the previous one
by the fact that one of the self-dual points \eqref{eq:selfdual}
remains finite while the other approaches infinity
\[
\tilde z^*_-=1,\qquad
\tilde z^*_+\sim 16\epsilon^{-4} .
\]
The parameters $a,b,c,d$ \eqref{eq:abcdclass}
for the supercharges in the fundamental representation remain finite
\[
a=\gamma,\qquad
b=-\frac{\tilde\alpha}{2\gamma}\,\frac{x-1}{x+1}\,,\qquad
c=\frac{2\gamma}{\tilde\alpha}\,\frac{1}{x-1}\,,\qquad
d=\frac{1}{\gamma}\,\frac{x}{x+1}\,,
\]
while the parameters $q,z$ are singular
and must be renormalised
\[
\tilde q=\epsilon q=-\coni\,\frac{x-1}{x+1}\,,\qquad
\tilde z=-4\epsilon^{-2} z=-\frac{4x}{(x-1)^2}\,.
\]
Using these parameters the fundamental 
r-matrix takes the same form as in \tabref{tab:rmatrix}.

For the algebra we have to rescale some generators 
\[
\tilde{\gencen}=\epsilon\gencen,\qquad
\tilde{\gender}=\epsilon^{-1}\gender.
\]
The parameters $U,V,W$ in \eqref{eq:matrices,eq:UVmatrix} 
for the Lie brackets \eqref{eq:DQmatrix,eq:DQaffder,eq:DQcentral} 
read in this case
\[
\tilde W=
\epsilon^{-1}W=
\conw \matr{cc}{-1&+\tilde\alpha\\-\tilde\alpha^{-1}\tilde z&+1},
\quad
U=
\frac{1}{4}\,\frac{\tilde z}{\tilde z-1}
\matr{cc}{-1&0\\-2\tilde\alpha^{-1}&+1},
\quad
V=-\frac{1}{2}\,\frac{\tilde z}{\tilde z-1}\,.
\]


\subsection{Special Trigonometric Case at \texorpdfstring{$h=0$}{h=0}}
\label{sec:trig0}

The limit $h\to 0$ was discussed already in \secref{sec:rational};
it yields the full rational r-matrix \cite{Klose:2006zd,Torrielli:2007mc,Beisert:2007ty}.
In this limit it was assumed that $x$ remains finite whereas $z\to 1$.
Likewise one can demand that $z$ remains finite and arbitrary
while $x\to 0$ or $x\to\infty$;
this turns out to yield an inequivalent limit. 
Let us consider the case of large $x$ 
\[
h=\epsilon,\qquad
z\sim \epsilon^0,\qquad
x=-i\epsilon^{-1}\frac{z-1}{z}+\order{\epsilon^0}.
\]
Then the parameters of the fundamental representation \eqref{eq:abcdclass,eq:zq} read
\[
a=\gamma,\qquad
b=c=0,\qquad
d=\frac{1}{\gamma}\,,\qquad
\tilde q=\epsilon^{-1}q=\coni\frac{1}{z-1}\,.
\]
This is almost the fundamental representation of the 
standard $\alg{gl}(2|2)$, but the central charge $\tilde q$ behaves differently.
Consequently, the r-matrix coefficients in \tabref{tab:rcoeffs} 
take a slightly non-standard form.
The case of $x\to 0$ leads to the conjugate fundamental representation.

Next, let us consider the algebra. 
In this case, we should rescale the generators $\gencen$ and $\gender$ 
according to 
\[
\tilde{\gender}=\epsilon\gender,\qquad
\tilde{\gencen}=\epsilon^{-1}\gencen
\]
in order to make their action finite. 
The algebra now takes the standard form 
with the parameters \eqref{eq:matrices,eq:UVmatrix}
\[
\tilde W=
\epsilon W=\conw (z-1)\matr{cc}{+1&0\\0&-1},
\qquad
U=0,
\qquad
V=-\frac{z}{z-1}\,.
\]
All the off-diagonal elements of the matrices are absent
as in the conventional affine $\alg{gl}(2|2)$. 
Only the central charge $\gencen$ appears with a non-trivial
dependence on the loop variable $z$.

In fact, we can formally make all the algebra relations 
like those for affine $\alg{gl}(2|2)$ by redefining 
the loop levels of $\tilde{\gencen}$ and $\tilde{\gender}$
\[
\bar{\gender}
=
(z-1)^{-1}\tilde{\gender},
\qquad
\bar{\gencen}
=
(z-1)\tilde{\gencen}.
\]
This leads to the standard affine algebra 
with parameters 
\[
\bar W=\conw \matr{cc}{+1&0\\0&-1},
\qquad
\bar U=\bar V=0.
\]
The simplification is however at the cost of changing the
universal r-matrix in \eqref{eq:reval}
because the transformation does not respect
the decomposition \eqref{eq:decompose}.

\subsection{Special Rational Case}
\label{sec:ratdef}

There is even a combination of the two different limits at $h\to 0$.
Here $h$ should approach $0$ faster than $z$ approaches $1$.
For example we can define the limit
\[
h=\epsilon^2,\qquad
z=1+i\epsilon u+\order{\epsilon^2},\qquad
x=\frac{u}{\epsilon}+\order{\epsilon^0}.
\]
The parameters of the fundamental representation reduce to 
\[
a=\gamma,\qquad
b=c=0,\qquad
d=\frac{1}{\gamma}\,,\qquad
\tilde q=\frac{1}{i\epsilon}\,q=-\coni\frac{1}{u}\,.
\]
The r-matrix diverges and becomes of rational type 
\eqref{eq:rrational}
\[\tilde r=i\epsilon r.
\]
The coefficients are almost those of the conventional rational
r-matrix, but there are a few important modifications
\[\begin{array}{c}
\displaystyle
\half(\tilde A_{12}+\tilde B_{12})
=\half(\tilde D_{12}+\tilde E_{12})
=\frac{\gamma_2}{\gamma_1}\,\tilde H_{12}
=\frac{\gamma_1}{\gamma_2}\,\tilde K_{12}
=\frac{1}{u_1-u_2}\,,
\qquad
\tilde C_{12}=\tilde F_{12}= 0,
\\[2ex]\displaystyle
\half(\tilde A_{12}-\tilde B_{12})=\half(\tilde D_{12}-\tilde E_{12})
=\frac{u_1-u_2}{4u_1u_2}\,,
\qquad
\tilde G_{12}=-\tilde L_{12}= \frac{u_1+u_2}{4u_1u_2}\,.
\end{array}
\]
The algebra is specified by the following parameters
\[
\tilde W=i\epsilon W=
\conw u\matr{cc}{-1&0\\0&+1},
\qquad
\tilde U=i\epsilon U=0,
\qquad
\tilde V=i\epsilon V=-\frac{1}{u}\,.
\]
This case may be viewed as the rational analog of the 
special trigonometric case at $h=0$ in \secref{sec:trig0}.


\begin{figure}\centering%
\includegraphics[width=\linewidth]{FigLimits.mps}%
\caption{Analytic structure and limits of various r-matrices.
T$(h)$: full trigonometric (\protect\secref{sec:classical});
T$(0)$: special trigonometric at $h=0$ (\protect\secref{sec:trig0});
T$(\infty)$: special trigonometric at $h=\infty$ (\protect\secref{sec:triginf});
T(twist): twisted trigonometric (\protect\secref{sec:trigtwist});
T(conv): conventional trigonometric (\protect\secref{sec:trigconv});
R(full): full rational (\protect\secref{sec:rational});
R(twist): twisted rational (\protect\secref{sec:rattwist});
R(def): special rational (\protect\secref{sec:ratdef});
R(conv): conventional rational (\protect\secref{sec:ratconv}).
Special points $z^\circ_\pm=0,\infty$ and $z^*_\pm$ are marked by 
$\circ$ and $\ast$, respectively.
A circle is drawn around coincident special points.
Two cases are connected by an arrow if the second is a particular
limit of the first.}
\label{fig:limits}
\end{figure}

\subsection{Summary}

In this section we have found more than a handful special limits 
of the $r$-matrix. What makes these limits special and 
how can we be sure that we have not missed an interesting case? 
To answer the question we should consider special points in the $z$-plane. 
The affine algebra specialises the two points $z^\circ_\pm=0,\infty$. 
Furthermore there are two points $z^*_\pm$ which 
lead to certain self-duality properties of representations,
see \eqref{eq:selfdual}.
In total there are four special points
\[
z^\circ_\pm=0,\infty,\qquad
z^*_\pm=(ih\pm h')^2.
\]

Above we have constructed limits by zooming into the neighbourhood of
certain points while potentially taking a simultaneous limit for $h$.
There is however a different point of view which makes the various limits
more transparent: By zooming into the neighbourhood of one point we 
effectively shift all other special points to the point at infinity. 
Hence the various limits correspond to grouping
the special points in different ways.

What is the role of the parameter $h$ in the limits?
Zooming into a neighbourhood can be achieved by M\"obius 
transformations of the $z$-plane with coefficients
depending on the limiting procedure. 
The transformation maps the special points to different positions, 
but there exist one conformal cross-ratio which remains invariant. 
Its value $s=(ih+h')^4$ is a function of $h$. 
Alternatively one can consider $h=h(s)$ to be a function of the cross-ratio $s$.
This allows us to view the four special points as independent,
and $h=h(z^\circ_\pm,z^\ast_\pm)$ as a function of their distribution 
modulo M\"obius transformations.

To understand the various limits, we should group the four special points
in all possible ways. Up to trivial permutations there are nine choices 
corresponding to the full trigonometric case with parameter $h$ and 
its eight limiting cases considered above,
see \figref{fig:limits}.
Note that the trigonometric cases have two distinct points
$z^\circ_\pm$ while the rational cases have identical points
$z^\circ_+=z^\circ_-$.
Two cases are linked by a limiting procedure 
if the special points of the first can be combined
to the special points of the second. 


\section{Conclusions and Outlook}
\label{sec:concl}

Classical r-matrices for Lie algebras were classified
in \cite{Belavin:1982aa}.
Three main classes,
distinguished by the distribution of poles in the complex plane,
were identified: rational, trigonometric and elliptic.
The classification is analogous for simple Lie superalgebras \cite{Leites:1984aa}.
In the case of the (non-simple) Lie superalgebra $\alg{gl}(2|2)$ 
an exceptional r-matrix was identified in \cite{Beisert:2007ty}.
This r-matrix is of rational type, but it is not of difference form. 
Its quantisation leads to Shastry's R-matrix for the Hubbard model \cite{Shastry:1986bb} 
or equivalently \cite{Beisert:2006qh} 
to the S-matrix for the AdS/CFT integrable system \cite{Beisert:2005tm}.
Hence this r-matrix is responsible for the exceptional integrable structure
in these models at the classical level. 

\smallskip

In this paper we have developed and investigated 
the trigonometric generalisation of the exceptional r-matrix for $\alg{gl}(2|2)$.
The corresponding fundamental quantum R-matrix was derived in \cite{Beisert:2008tw},
and it defines the integrable structure of the Alcaraz--Bariev model \cite{Alcaraz:1999aa}
(type B).
As for the rational case, the underlying Lie algebra is a deformation 
of the loop algebra $\alg{gl}(2|2)[z,z^{-1}]$. 
The deformation is special in the sense that 
the Lie brackets are not homogeneous in the level of the loop algebra. 
Nevertheless, the algebra admits solutions to the classical Yang--Baxter equation.
Analogously, it admits a decomposition into positive and negative subalgebras.
An interesting feature of the algebra is 
that it has one modulus $h$ 
whose value has significant impact on the algebra.
One may wonder whether there are other similar cases
of deformed loop algebras or if $\alg{gl}(2|2)$ is truly
exceptional in this regard.
In other words, which is the precise (co)homological 
property of $\alg{gl}(2|2)$ or its loop algebra
giving rise to the deformation?

\smallskip

The deformed loop algebra also admits the extension 
by a derivation and a central charge to an affine Lie algebra.
This algebra is not of Kac--Moody type, 
but its structure is similar in many respects.
The affine derivation serves as a scaling of the loop variable $z$ 
(or a shift in $u$ in the rational case). 
In a physical scattering context, it can be viewed as
a boost operator in analogy to Lorentz boosts in two spacetime dimensions. 
Also we must extend the notion of particles to fields,
because the particle momentum does not commute with boosts.
Interestingly, the boost has non-trivial cobrackets,
hence the symmetry should be viewed as deformed or non-commutative \cite{Gomez:2007zr,Young:2007wd}.
Non-invariance of the r-matrix also explains the violation 
of difference form for the r-matrix.
Finally, extension of a symmetry often leads to additional restrictions.
Here it would be interesting to see if, e.g.,
the overall prefactor of the r-matrix can be constrained
by the affine extension.

\smallskip

Subsequently, we have investigated discrete transformations
and special points of the r-matrix. 
Transformations include conjugation, 
inversion of the loop variable, a flip of statistics and
a duality for the global parameter. 
Conjugation maps different representations into each other.
In particular, the family of fundamental representations is self-conjugate, 
and thus conjugation extends to a crossing symmetry of the r-matrix,
cf.\ \cite{Janik:2006dc}.
Inversion symmetry of the r-matrix can be viewed 
as a scattering unitarity condition. 
The statistics flip interchanges bosons and fermions in the fundamental representation. 
At the level of the algebra it permutes the two $\alg{sl}(2)$ subalgebras.
Last but not least, the duality map relates algebras/r-matrices
with different moduli $h$. 
An important insight gained from the discrete transformations
is that next to the special points $z=z^\circ_\pm=0,\infty$, 
which exist for any trigonometric r-matrix, 
there are two self-dual points $z=z^*_\pm$ whose value depends on $h$.

\smallskip

Finally, several r-matrices with simpler structures 
were recovered as limiting cases. 
For example, our trigonometric r-matrix reduces 
to the exceptional rational r-matrix of \cite{Klose:2006zd,Torrielli:2007mc,Beisert:2007ty}
in a particular limit. The latter can be reduced further to the conventional
rational $\alg{gl}(2|2)$ r-matrix as well as to two other intermediate cases.
In total there is the one-parameter family 
of exceptional trigonometric r-matrices and 8 singular cases,
see \figref{fig:limits}.
The trigonometric family has the most sophisticated structure
while the conventional rational r-matrix is the plainest:
All intermediate cases can be obtained from the former and be reduced to the latter.
These include some special cases with $\alg{gl}(2|2)$ structure 
discovered earlier in various contexts:
They can be of trigonometric or of rational type,
they are conventional or deformed and
untwisted or $\Integers_2$-twisted.
In terms of algebra all cases follow from the one discussed in this paper:
Its structure can be simplified through limits and algebraic contractions 
down to the plain $\alg{gl}(2|2)$ affine Kac--Moody algebra.
It would be interesting to find out whether
the trigonometric structure is itself a limiting case
of some exceptional elliptic r-matrix
(note that both $\alg{psl}(2|2)$ and $\alg{osp}(4|2)$ admit
elliptic r-matrices \cite{Leites:1984aa}).

\medskip

With a good part of the classical framework established, 
several open questions concerning the
exceptional trigonometric r-matrix remain. 
For instance, we would like to promote the Lie bialgebra 
to a quantum affine Hopf algebra (cf.\ \cite{Etingof:1995aa}). 
Are there any obstacles
due to the non-standard structure of the affine algebra?
So far only the fundamental quantum R-matrix has been established. 
However there is little doubt that R-matrices for 
higher representations can indeed be constructed 
as in the rational case \cite{Beisert:2006qh,Chen:2006gp,Arutyunov:2008zt,deLeeuw:2008dp,deLeeuw:2008ye,Arutyunov:2009mi,Arutyunov:2009ce,Arutyunov:2009iq,Arutyunov:2009pw}.
This would be very suggestive of a universal R-matrix.%
\footnote{Doubts raised in \protect\cite{Arutyunov:2009pw}
apply only to a different type of quantum algebra without derivations $\gender$
and with a minimal set of Serre relations.} 

\smallskip

Developing the quantum affine algebra would
establish, as a by-product, the Yangian for the undeformed 
Hubbard model or for integrable scattering in AdS/CFT.
One complication in the formulation might reside in the existence 
of the tower of derivations $z^n\gender$ 
for which Drinfeld's first presentation \cite{Drinfeld:1985rx,Drinfeld:1986in}
using Chevalley--Serre generators is not ideally suited.
Instead, Drinfeld's second realisation \cite{Drinfeld:1988aa}
along the lines of \cite{Spill:2008tp} may prove to be more helpful.

\begin{figure}\centering
\includegraphics{FigDynkinOX-O.mps}
\caption{Dynkin diagram for $\alg{d}(2,1;0)$.}
\label{fig:Dynkin2}
\end{figure}

\begin{figure}\centering
\includegraphics{FigDynkinOXO.mps}
\qquad
\includegraphics{FigDynkinXXX.mps}
\qquad
\includegraphics{FigDynkinXOX.mps}
\caption{All Dynkin diagrams for $\alg{sl}(2|2)$.}
\label{fig:Dynkin3}
\end{figure}

Also the choice of Dynkin diagram may play a role:
For instance, the Bethe equations \cite{Lieb:1968aa,Beisert:2005fw} 
cannot be formulated (easily) 
for the distinguished diagram in \figref{fig:Dynkin}
(leftmost in \figref{fig:Dynkin3}),
but it appears to prefer a structure 
reminiscent of the exceptional superalgebra 
$\alg{d}(2,1;\alpha)$ with singular parameter $\alpha=0$ in \figref{fig:Dynkin2}.
The latter has a non-symmetrisable Cartan matrix, cf.\ \cite{Hoyt:2007aa}.
It would be interesting to derive the r-matrices 
for the various other Dynkin diagrams (see \figref{fig:Dynkin3}),
and to understand how to transform between them,
see also \cite{Khoroshkin:1994aa,Geer:2005aa}.

\paragraph{Acknowledgements.}

The author thanks
B.\ Hoare, 
T.\ McLoughlin,
V.\ Schomerus,
V.\ Serganova,
M.\ Staudacher and
A.\ Tseytlin
for interesting discussions.
Useful comments on the manuscript by referees are acknowledged.
The author acknowledges hospitality 
by the Galileo Galilei Institute and Durham University 
during the workshops ``Non-Perturbative Methods in Strongly Coupled Gauge Theories'' (GGI),
``New Perspectives in String Theory'' (GGI) and ``Gauge and String Amplitudes'' (Durham)
where part of the present work was performed.

\appendix

\pdfbookmark[1]{\refname}{references}
\bibliography{classtrig}
\bibliographystyle{nb}

\end{document}